# Nanostructuring Semiconductor Surfaces Using Self-Assembled Metal Nanoparticles for Optoelectronic Applications

S. Catalán-Gomez[1,2]*, M. Ibáñez[1], P. Guo[3], V. Braza[4], B. Vázquez[1], R. Sun[3], F. Li[3], P. Ibáñez[1], Y.K. Ryu[1], M. Vila[5], J. Bartolomé[5], D. F. Reyes[4], J. Martínez[1], D. González[4], H. Wang[3], J.M. Ulloa[1] and A. Hierro[1]

[1]Instituto de Sistemas Optoelectrónicos y Microtecnología, Universidad Politécnica de Madrid, ETSI Telecomunicación, Av. Complutense 30, E-28040 Madrid, Spain

[2]Departamento de Ingeniería Eléctrica, Electrónica Automática y Física Aplicada, Universidad Politécnica de Madrid, ETSI y Diseño Industrial, Ronda de Valencia 3, E-28012 Madrid, Spain

[3]State Key Laboratory of Solidification Processing, Center for Nano Energy Materials, Northwestern Polytechnical University, Xi'an, P. R. China.

[4]Departamento de Ciencia de los Materiales e Ingeniería Metalúrgica y Química Inorgánica, Universidad de Cádiz, E-11510 Puerto Real, Cádiz, Spain

[5]Escuela de Ingeniería de Fuenlabrada, Universidad Rey Juan Carlos, Fuenlabrada, E-28942 Madrid, Spain

*Corresponding author:

sergio.catalan.gomez@upm.es

## Abstract

Semiconductor nanostructures are key light-management elements because their geometry can strengthen light-matter interaction, suppress reflection, and enhance optical coupling. Yet their adoption across material platforms is constrained by lithography, multistep pattern transfer, persistent masks, and material-specific etching. Here, we introduce a lithography-free two-step strategy in which thermally evaporated self-assembled Ga nanoparticles act as self-consuming masks during Ar inductively coupled plasma reactive-ion etching. Correlated transmission electron microscopy on GaAs reveals the formation mechanism. Comparable erosion of the Ga nanoparticles and exposed substrate progressively converts Ga-capped pedestals into pure semiconductor nanocones while preserving crystalline order. Nanocone dimensions extend from tens to hundreds of nanometres, and surface coverage is tunable through Ga deposition and plasma power. The same process produces nanocones on GaP and Si and dense nanostructures on polycrystalline hybrid metal-halide perovskite without detectable $PbI_2$ formation. During nanostructuring, the sun-weighted reflectance reaches minima of 11-15% on the monocrystalline substrates, whereas the perovskite achieves 0.5% after complete mask consumption. The approach therefore combines a self-assembled mask, its consumption, and pattern transfer in a single physical etching step, providing a simple route to light-coupling surfaces for III-V, Si, and perovskite optoelectronics.

Semiconductor nanostructures provide a powerful means of engineering optical functionality beyond that available in planar materials. By tailoring geometry at subwavelength scales, they enable control over light coupling, scattering, and confinement, making surface architecture an additional design parameter alongside material composition. Tapered nanostructures are particularly attractive because they can provide a gradual effective refractive-index transition between air and the semiconductor, suppressing reflection while promoting light trapping and absorption[1]. These functions make semiconductor nanostructures attractive for solar cells[1], photodetectors, light emitters and integrated photonics[2], but practical deployment requires fabrication methods that combine nanoscale control, large-area scalability, low damage, and compatibility with very different semiconductor chemistries.

Established fabrication routes rarely meet all these requirements simultaneously. Bottom-up epitaxy enables high-quality crystalline growth but requires carefully controlled growth conditions and, for heterogeneous integration, strategies to accommodate lattice and thermal-expansion mismatch[3]. Conventional top-down patterning offers positional precision but combines lithography, resist processing, pattern transfer, and mask removal. For metal-halide perovskites, ionic lattices and solvent sensitivity further complicate lithographic processing, requiring adapted procedures[4] even in successful demonstrations of device integration[2].

Self-assembled masks provide a scalable alternative, with nanosphere lithography being the most established example[5,6]. It offers a low-cost route to large-area patterning and has been used to fabricate antireflective GaAs nanostructures[7] and templates for perovskite nanopillar arrays[8]. Its process flow commonly involves colloidal assembly, plasma-mediated mask modification, substrate etching and removal of the remaining mask, although these operations can be integrated into a continuous plasma process[6,9,10]. Sphere packing can also generate voids, multilayers, and nonuniform domains, whereas the dimensions and arrangement of the final features remain linked to the starting colloids[5,9]. Metal-assisted chemical etching offers another powerful route to high-aspect-ratio semiconductor structures, including III-V semiconductors[11,12], but depends on a metal catalyst, oxidants, wet chemistry, and often a predefined catalyst pattern. Its mechanism and morphology are strongly coupled to catalyst geometry, crystal orientation, and reaction conditions[13,14].

The central challenge is therefore to create optically functional semiconductor nanostructures without lithographic patterning, permanent hard masks, reactive halogen chemistry, or a separate mask-removal step. A self-assembled mask that disappears while transferring its geometry would substantially simplify the process and could improve compatibility with chemically fragile absorbers. Gallium nanoparticles are well suited to this role. Thermal evaporation can produce liquid Ga nanodroplets directly on diverse substrates[15–18] and exposure to air creates a self-limiting native oxide shell around the metallic core[19].

Here, we demonstrate semiconductor nanocone formation using self-assembled Ga nanoparticles as self-consuming masks during Ar ICP-RIE. Cross-sectional microscopy on GaAs resolves the coupled erosion of the mask and substrate, while experiments on GaP, Si and polycrystalline perovskite establish the transferability of the approach. We examine how deposition and plasma conditions control nanocone geometry and coverage and evaluate the resulting structural and optical properties.

## GaAs nanocone formation through Ga-mask consumption

Before etching, Ga spontaneously forms hemispherical NPs on GaAs upon thermal evaporation (Fig. 1a). This behaviour is well documented for Ga[15] not only in this substrate but also on a variety of substrates and is due to its high surface tension and poor wettability, which favors droplet formation instead of continuous films[17]. The annular dark-field (ADF) STEM overview reveals an array of Ga-NPs with sizes about dozens of nm. High-magnification ADF image (Fig. 1 a2) show their hemispherical shape, in agreement with previous reports, while the FFT (inset) shows only a diffuse halo, consistent with a non-crystalline (liquid) Ga core enveloped by a thin amorphous shell. Energy-dispersive X-ray (EDX) mapping in Fig. 1 a3) identifies the core-shell structure being the latter a self-limited native $Ga_xO_y$ layer with a thickness of about 2-3 nm, which passivates the liquid Ga core and protects it from the environment as well studied before[19].

When exposed to the plasma, high-energy Ar ions bombard the exposed GaAs regions between adjacent Ga-NPs, thereby initiating the substrate etching process. Simultaneously, the Ga-NPs undergo gradual erosion, which is likely to involve the removal of the native oxide shell and the underlying metallic core. Specifically, as exemplified in Fig. 1 b1) for 50 s of etching time, the areas between Ga-NPs recede, while each NP remains situated on a slightly elevated GaAs pedestal. These "hybrid nanostructures" consist of a metallic cap supported by a truncated semiconductor cone as observed in the EDX mapping (Fig. 1 b3). As the process continues to 100 s (Fig. 1c), the vertical contrast increases as the GaAs is further removed from the open and non-protected areas. The narrowing of the cones is indicative of the progressive reduction in droplet size consequent to Ga consumption. The FFT pattern on the GaAs pedestal exhibits sharp diffraction spots corresponding to the GaAs lattice, confirming that the crystalline structure of the substrate is preserved during etching. High-resolution ADF-STEM imaging (Figure S1) further shows that the GaAs lattice extends up to a thin amorphous surface layer on both the exposed pedestals and the side facets of the fully formed nanocones. This layer, of approximately [3.6 ± 0.3] nm, is comparable to or thinner than the native amorphous interfacial layer beneath unetched Ga-NPs (approximately [4.9 ± 0.5] nm) and is consistent with the native oxide formed upon air exposure. No additional amorphized GaAs is resolved beneath it, indicating that any plasma-induced damage is confined to the outermost few nanometres. EDX analysis (Fig. 1 c3) reveals a significantly reduced Ga signal in the cap, while the underlying cone shows a stoichiometric Ga:As ratio. By 150 seconds (Fig. 1d), the Ga cap from the Ga-NP at the apex disappears, indicating the complete consumption of the mask and resulting in well-defined, pure GaAs nanocones. These GaAs nanocones maintain epitaxial continuity with the bulk substrate, as corroborated by the corresponding FFT patterns in Fig. 1 d2) and by lattice fringes extending to the nanocone facets (Fig. S1 e)). Raman measurements (Figure S2) on flat GaAs and the GaAs nanocones shown in Fig. 1d further support the preservation of crystalline order: the TO phonon remains at 268 ± 0.5 cm$^{-1}$, while its intensity increases following nanostructuring. This enhancement is consistent with the modified scattering geometry introduced by the inclined nanocone surfaces, which can enable TO scattering that is forbidden in ideal backscattering from the flat (001) surface[20].

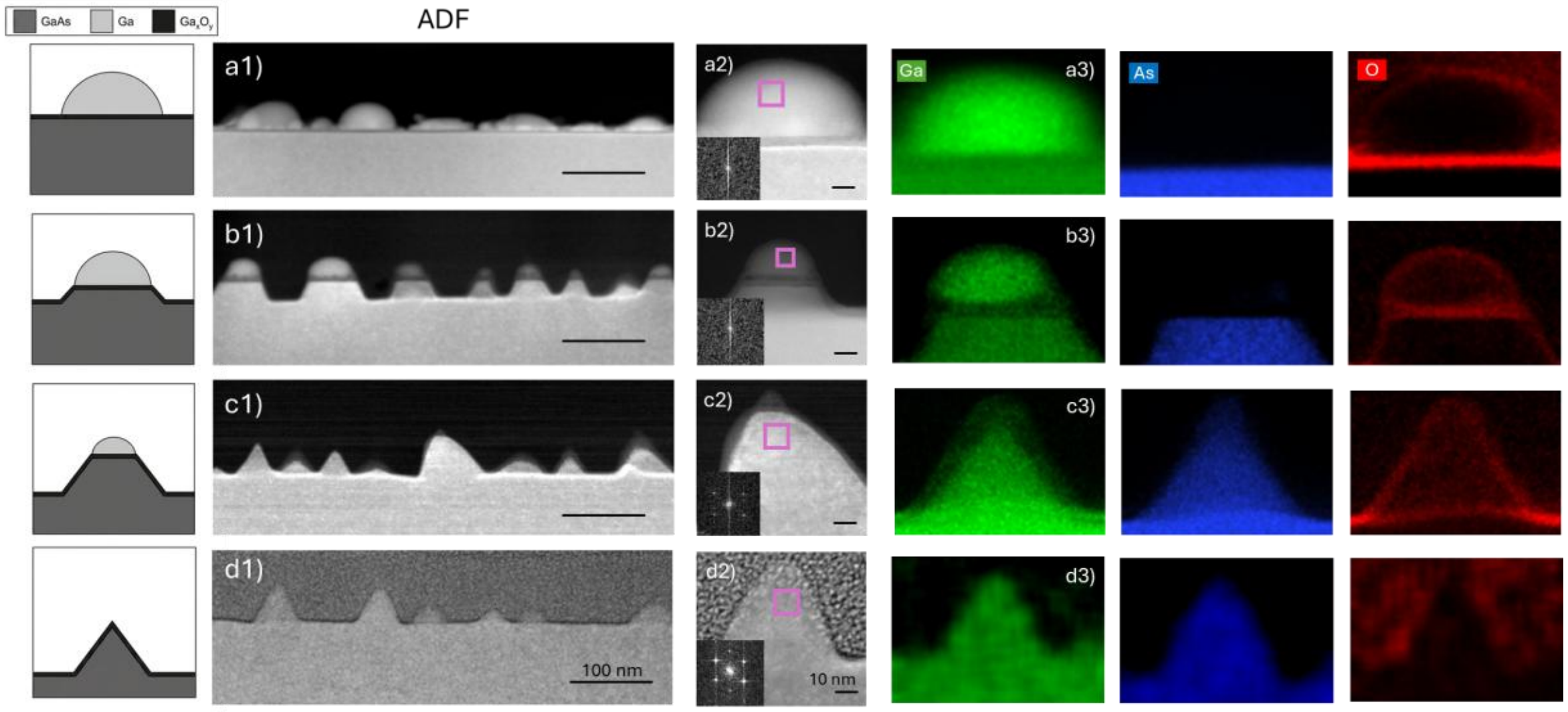

*Figure 1. Self-consuming Ga nanoparticles govers the formation of epitaxial GaAs nanocones during Ar ICP-RIE. Schematic illustration (first column) and corresponding cross-sectional STEM micrographs (second column) depicting the ICP-RIE process using Ga-NPs as a self-consuming mask. The time series tracks the structural transformation at (a) 0 s, (b) 50 s, (c) 100 s, and (d) 150 s of etching time. High-magnification STEM micrographs including FFT patterns (third column), and EDX maps of Ga, As and O (fourth to sixth column) for each specific time.*

A key aspect of this self-consuming behaviour is that the etch rates of Ga and GaAs under our Ar ICP-RIE conditions are of the same order of magnitude. This balance ensures that the Ga mask is neither etched much faster than the substrate (which would do not produce any significant pattern) nor so slowly that it survives as a residual metallic layer. Instead, Ga and GaAs are removed concurrently: the Ga-NP shrinks while the surrounding semiconductor recesses, naturally sculpting a cone whose height increases as the process proceeds, until the metal has fully vanished. The quantitative calibration of these etch rates, obtained by combining AFM step-height measurements on GaAs reference samples and TEM data is provided in Supplementary Note 1 and Figure S3.

## Control of nanocone dimensions and coverage

Figure 2 a) compares the AFM topographies of small (30 s evaporation time) and large (150 s evaporation time) self-assembled Ga-NPs before etching and the corresponding GaAs nanocones after complete mask consumption. Nanocone formation is achieved after 150 s for the smaller nanoparticles and 420 s for the larger ones, demonstrating that the process operates across different initial feature sizes. The corresponding height profiles in Fig. 2 b) quantify the accompanying shape evolution. For the smaller nanostructures, the height-to-base-diameter ratio increases from 0.44 ± 0.02 for the initial Ga-NPs to 0.75 ± 0.05 for the GaAs nanocones. The larger structures exhibit a similar transformation, with the aspect ratio increasing from 0.51 ± 0.04 to 0.8 ± 0.1 and final nanocone heights reaching approximately 320 nm. These results show that increasing the initial NP size produces larger nanocones while retaining a comparable final aspect ratio.

The transformation of the larger Ga-NPs, with an initial mean radius of approximately 140 nm, is further resolved by the time-dependent top-view and cross-sectional SEM measurements in Supplementary Fig. S5. As for the smaller NPs examined in Fig. 1, intermediate Ga-capped GaAs pedestals evolve into semiconductor nanocones as the metallic mask is consumed. The Ga and GaAs etching rates extracted from these images agree with the calibration presented in Supplementary Note 1, supporting the same concurrent mask-consumption and substrate-etching mechanism over the investigated size range.

Nanocone formation also reduces the surface coverage, as evident from the sparser distributions after etching in Fig. 2 a). For the smaller nanoparticles, the complete AFM time series in Supplementary Note 2 and Fig. S4 quantifies a decrease from approximately 85% initial Ga coverage to approximately 28% GaAs nanocone coverage after 150 s, consistent with the complete mask consumption observed by TEM. This evolution accompanies the progressive lateral shrinkage of the Ga mask and recession of the exposed substrate.

Beyond the changes associated with etching time, the final coverage can be adjusted through the plasma conditions. Figure 2 c) compares arrays obtained from the same initial Ga-NP distribution using different ICP powers, while Fig. 2 d) summarizes the corresponding coverage changes. At 100 W, complete mask consumption after 7 min yields a nanocone coverage of approximately 25%. Increasing the ICP power to 400 W shortens the required etching time to 5 min and produces a substantially sparser array with approximately 5% coverage. These results demonstrate that plasma power provides a means of tuning the final nanocone coverage within the same two-step fabrication process, without additional patterning or mask-removal steps.

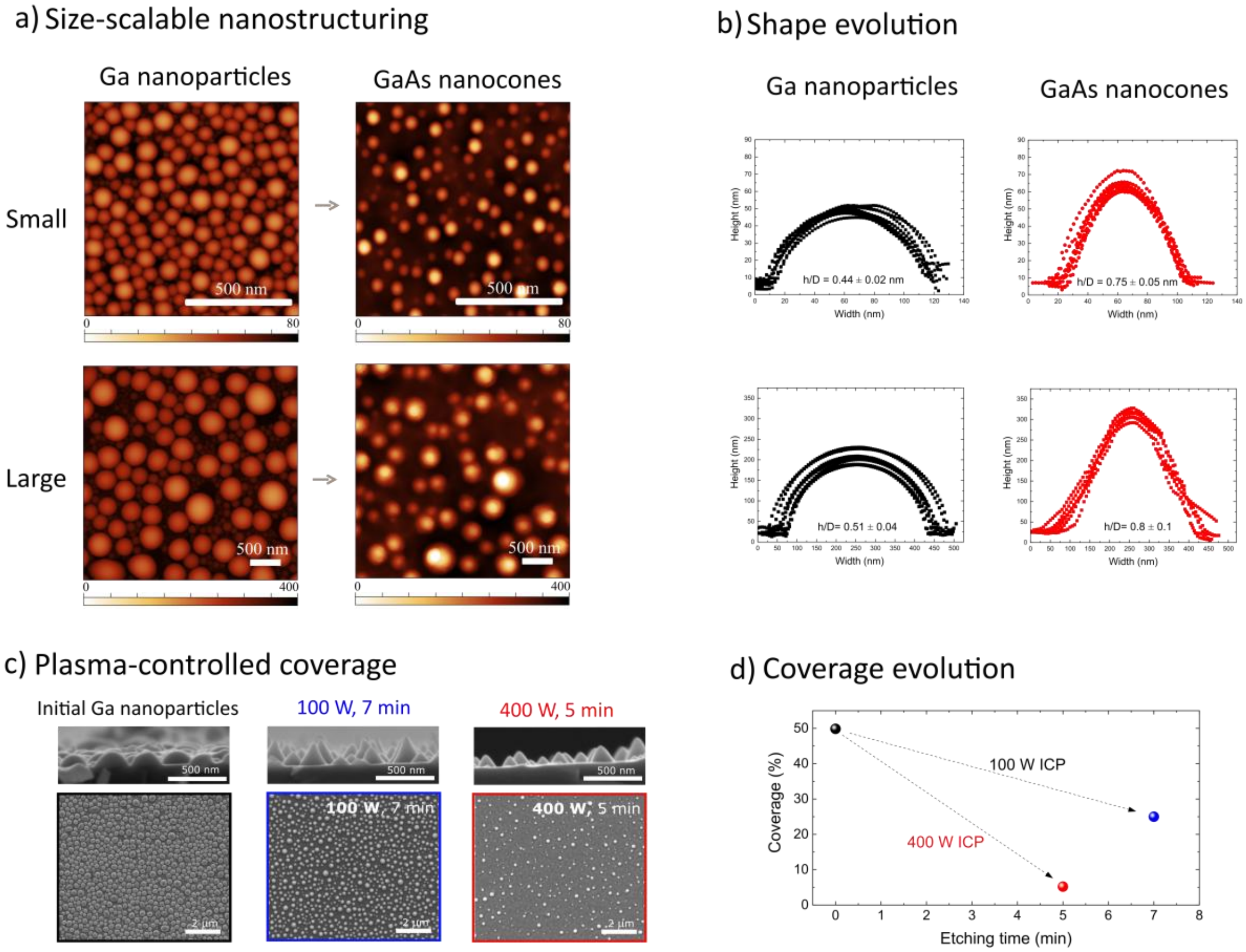

*Figure 2. Size scalability, shape evolution and coverage control of Ga-NP-assisted nanostructuring on GaAs. (a) AFM height images of small (top row) and large (bottom row) self-assembled Ga-NPs before etching and the corresponding GaAs nanocones after 150 and 420 s of Ar ICP-RIE, respectively. (b) Representative height profiles corresponding to the nanostructures in (a). The aspect ratio is expressed as h/D, where h is the height and D is the base diameter. (c) Cross-sectional and top-view SEM images of the initial large Ga-NPs and the resulting GaAs nanocone arrays obtained at ICP powers of 100 W for 7 min and 400 W for 5 min. (d) Initial and final surface coverage for the two ICP powers. Dashed arrows serve as guides to the eye.*

## Transfer to GaP and silicon

We have applied an identical protocol (Ga-NP deposition by thermal evaporation followed by Ar ICP-RIE under the standard conditions calibrated for GaAs) to two additional III-V and group-IV substrates: gallium phosphide (GaP) and Si. The detailed transformation series for both materials with multiple intermediate etching times showing hybrid nanostructures analogous to the GaAs case are provided in Supplementary Figure S6 a) for GaP and S7 a) for the Si case.

Atop GaP, the as-deposited Ga-NPs exhibit a bimodal size distribution (Fig. S6 b), very similar to the large Ga-NP ensemble used in GaAs (Fig. 2 a), bottom row). Specifically, Ga-NPs with radius of around 112 ± 5 nm are formed surrounded by smaller ones. After etching under the standard Ar ICP-RIE conditions, top-view SEM image (Fig. 3 b) and corresponding size histogram (Fig. S6 c) reveal well-defined GaP nanocones with relatively low size dispersion, base radii reduced to ≈79 ± 6 nm and surface coverage of ≈25%, comparable to the GaAs case. Cross-sectional SEM images show well-defined GaP nanocones with no discernible apical caps. This morphological evolution is consistent with consumption of the Ga-NP mask, although SEM imaging alone does not establish the absence of residual Ga.

An analogous behavior is observed on Si (Fig. 3 d) and Supplementary Fig. S7). The initial Ga-NPs again show a bimodal distribution with Ga-NPs radius peaked around 79 ± 3 nm surrounded by smaller ones (Fig. S7 b), demonstrating that Ga self-assembly proceeds similarly on this group-IV substrate in agreement with previous studies[18]. After the Ar ICP‐RIE procedure, top-view, the cross-section SEM image and corresponding histogram (Fig. 3 e) and Fig. S7 c)) display an array of Si nanocones whose base radii shrink to ≈40 ± 2 nm and a coverage of ∼12%. In both materials, the removable Ga-NP mask dictates the pattern transfer, and the step-by-step evolution from Ga-NPs to pure semiconductor nanocones proceeds in the same way regardless of the underlying semiconductor chemistry or lattice constant.

To assess whether nanostructuring preserves the crystalline integrity of the host materials, we performed Raman spectroscopy on GaP and Si before and after Ar ICP-RIE (Fig. 3 c,f). For GaP, the longitudinal-optical (LO) and transverse-optical (TO) phonon modes near 405 and 365 $cm^{-1}$, respectively, retain their positions after nanocone formation, with FWHM changes of approximately 0.1 $cm^{-1}$, below the experimental uncertainty of 0.5 $cm^{-1}$. The increased TO-to-LO intensity ratio is consistent with the modified scattering geometry introduced by the inclined nanocone surfaces, which can enable TO scattering that is symmetry-forbidden in ideal backscattering from the flat (001) surface[20]. The reduced contribution of the original planar surface may also contribute to this redistribution of Raman intensity. These intensity changes, together with the unchanged peak positions and linewidths, are consistent with morphology-dependent Raman scattering without detectable disruption of crystalline order. For Si, the optical phonon near 520 $cm^{-1}$ likewise retains its position within experimental uncertainty and its narrow linewidth, supporting the preservation of crystalline order following plasma exposure.

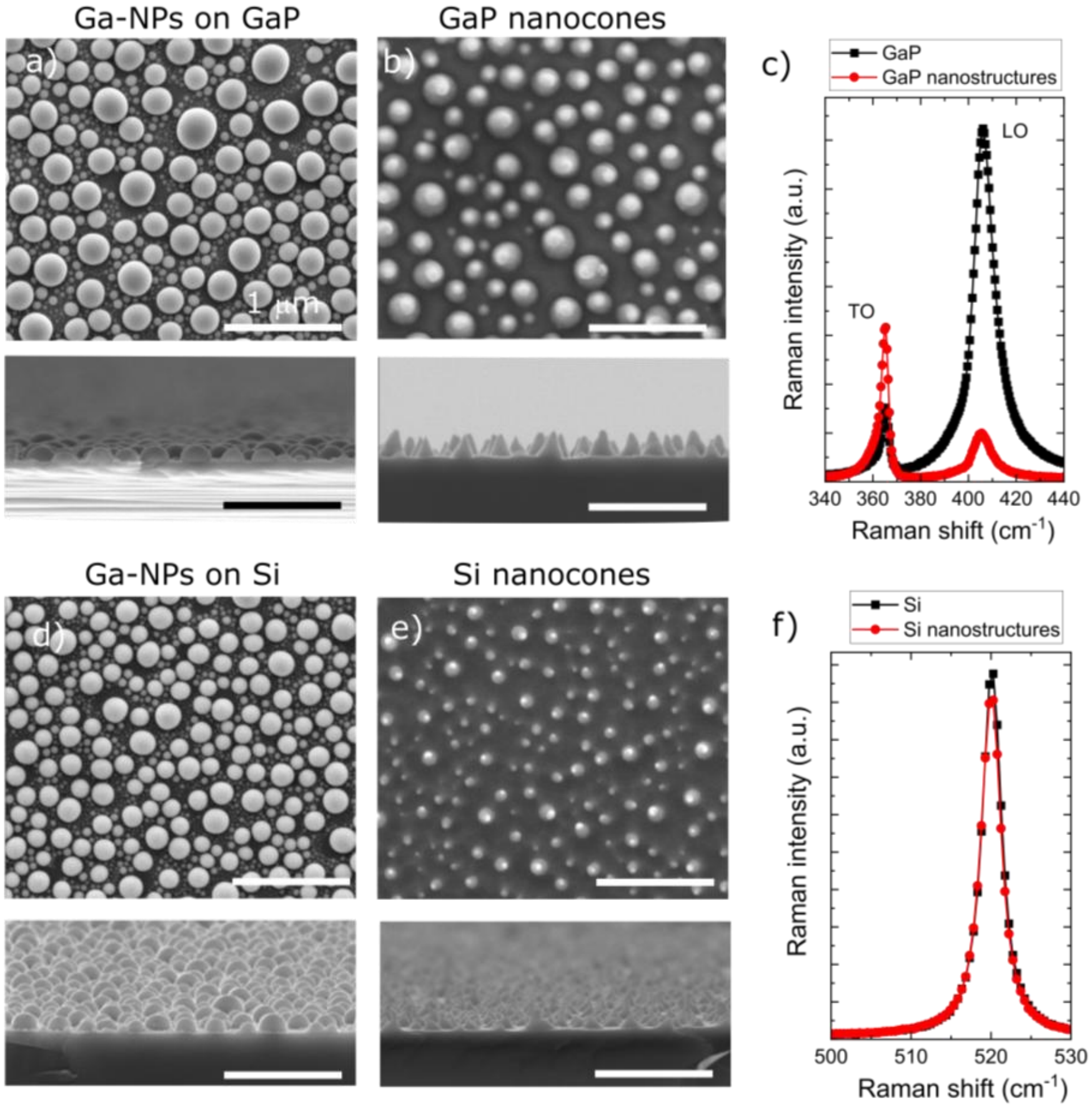

*Figure 3. Extension of Ga-NP-assisted nanocone formation to GaP and Si. (a,d) Top-view and cross-sectional SEM images of the initial self-assembled Ga-NP distributions on plain GaP and Si substrates, respectively. (b,e) SEM images of the corresponding GaP and Si nanocone arrays after Ar ICP-RIE of a) and d). (c,f) Raman spectra of the pristine and nanostructured GaP and Si substrates, respectively. All scale bars corresponds to 1µm.*

## Nanostructuring of hybrid perovskites

$Cs_{0.05}FA_{0.85}MA_{0.1}PbI_3$ films prepared from 2 M precursors had an initial thickness of approximately 1.1 µm and an average grain size of 0.75 µm. Bare control films and Ga-NP-coated target films were subjected to identical Ar ICP-RIE treatments for up to 5 min to compare nanostructure formation and material preservation as illustrated in Figure 4 first and second row. Notably, the Ga deposition on perovskite (Figure 4 b1)) produces a size distribution that is less sharply bimodal than on monocrystalline substrates as shown in SEM images with different magnification (Figure S8 and the corresponding histograms likely due to the lower Ga diffusion across polycrystalline grain boundaries and poorer wetting, higher contact angle, visible in the cross-sectional SEM image of Fig. S8 e). On the target surface, SEM top-views show the characteristic progression: initial Ga-NPs give way step by step to sharp, directionally etched nanocones whose density, spacing and overall array highly improve those obtained on GaAs, GaP and Si. In direct contrast, the control films exhibit no such nanostructuring. Short etching times (<2 min) induce only mild, isotropic roughening with grain boundary blurring, while longer exposures produce irregular, low-contrast features distributed homogeneously across the sample (see Figure S9 the same SEM study but with low-magnification images).

Cross-sectional SEM of the target perovskite (Fig. 4c) confirms the formation of well-defined nanocones and a gradual thickness reduction of the underlying film during Ar ICP-RIE. A dedicated thickness calibration based on cross-sectional SEM measurements at different positions across 1.5 × 1.5 cm samples (Figure S10) yields a perovskite etch rate of $1.9 \pm 0.3\ nm \cdot s^{-1}$ which is higher than monocrystalline semiconductors in the previous section consistent with their softer crystalline structure. The relatively large error bars reflect thickness variations intrinsic to spin-coated films rather than fluctuations in the plasma process itself. This rate, despite being larger than for the monocrystalline materials reported before, indicates that both materials are removed on comparable time scales, ensuring that the transient Ga-NP mask persists just long enough for nanocone formation before being fully self-consumed.

AFM provides quantitative depth to the contrast impact of the plasma etching in both type of substrates as shown in Figure S11. The pristine film's AFM image (Fig. S11 a)) shows a RMS roughness of 22 nm typical of solution-processed perovskites that only increases to a mere 27 nm in the control after 5 min (Fig. S11 b)). The target (Fig. S11 c)), however, yields to 77 nm rms, dominated by the nanocone structures. High-resolution 1×1 $\mu m^2$ topography and their horizontal lines and profiles (Fig. S11 d) to f)) unveils individual nanocones with base radius ∼50-80 nm, heights exceeding their widths (aspect ratio ∼1.25) and smooth facets, confirming the similar self-consuming physics mechanism operative across all substrates.

In order to address the chemical integrity, Raman spectroscopy characterization was used to evaluate the possible degradation produced by the implemented etching procedures in the final etched control and target samples together with the as-grown perovskite film. The representative spectra for each case are shown in Fig. 4 d) while the microscope corresponding images are included in Fig. S12 a). The target spectrum closely tracks the as-deposited reference, retaining broad organics-cation methylammonium (MA) vibrational features at ∼350 and 440 $cm^{-1}$[21] with negligible evolution, with no emergence of the $PbI_2$ marker at 111.7 $cm^{-1}$ (see Raman spectra of a pure $PbI_2$ crystal in Fig. S12 b). The control, by contrast, displays $PbI_2$ peak growth (intensity five-fold) alongside suppression of perovskite signatures, indicating severe surface degradation[22]. It is worth noting that the Raman measurement acquisition time was pre-calibrated in the as-grown perovskite film before etching to prevent laser-induced artefacts as shown in the microscope images of Fig. S12 c). Those conditions were applied to the control and target etched samples.

Low-temperature (10 K) photoluminescence (PL) spectra of the as-grown control, etched control and Ga-NP-nanostructured target perovskite are shown in Fig. 4 e). The target exhibits approximately 55-fold higher PL peak intensity than the as-grown control, whereas direct plasma etching of the bare film reduces it by a factor of approximately six. The corresponding full widths at half maximum (FWHM) are 48.6, 47.5 and 10.5 meV for the as-grown control, etched control and target, respectively. The enhanced and narrower emission is consistent with reduced non-radiative losses and lower energetic heterogeneity in the emitting regions. Since the nanocone lateral dimensions (<300 nm) are smaller than the average pristine grain size (≈750 nm), some nanocones may lie within individual grains, potentially reducing the influence of grain-boundary-associated recombination[23,24]. However, changes in excitation absorption and light extraction may also contribute to the PL enhancement.

Finally, the optical viability of the nanostructured perovskite for photovoltaics hinges on preserving its light-harvesting capability. To assess this, we measured the absorbance spectra of both control and target samples (Figure S13 a) and b) and quantified the evolution by computing the sun-weighted absorbance (SWA) according to equation (1):

$$SWA = \frac{\int S(\lambda)\cdot A(\lambda)\cdot d\lambda}{\int S(\lambda)\cdot d\lambda} \qquad \text{Equation 1}$$

Here, $A\ (\lambda)$ is the absorbance spectrum and $S\ (\lambda)$ is the AM1.5G solar photon flux. Figure 4 f) shows the evolution of SWA over 300-800 nm, normalized to the pristine, untreated perovskite control. For the directly etched control, SWA decreases monotonically, reaching approximately 87% of the reference value after 5 min. This decrease is consistent with film thickness reduction during plasma etching (Fig. S10), although chemical degradation may also contribute. In contrast, the target exhibits an initial increase in SWA following Ga-NP deposition, potentially associated with the modified optical response introduced by the metallic NPs, whose localized surface plasmon resonances lie within the measured spectral range[18,25]. After 5 min of etching, the target retains approximately 94% of the reference SWA.

A closer inspection of the absorbance spectrum (Fig. S13) reveals that the spectral evolution differs markedly between control and target. In both cases, the apparent bandgap position around 794 ± 2 nm extracted from the derivative of the spectra remains essentially unchanged with etching time, indicating that the fundamental optical transition of the perovskite is preserved. However, in the target, the entire spectrum decreases almost uniformly with etching time, consistent with a simple reduction of the effective thickness of an otherwise intact perovskite layer. In the control, by contrast, the absorbance in the 600-800 nm range (near the perovskite band edge) drops strongly, whereas the 300-500 nm region remains nearly constant. This behaviour is characteristic of the growing contribution of $PbI_2$, whose bandgap is $\sim$2.4 eV[26] and predominantly absorbs in this high-energy range, and thus provides optical confirmation of its formation in agreement with the previous Raman analysis. For the target sample, the Urbach energy (Fig. S13 c)) remains essentially constant within experimental uncertainty, indicating that the electronic disorder and defect density of the perovskite are not significantly altered by the Ga-NP-assisted etching[27]. In the control, however, the Urbach energy increases noticeably for etching times longer than 3 min, suggesting an accumulation of sub-gap states and enhanced disorder consistent with the combined morphological and spectroscopic signatures discussed above.

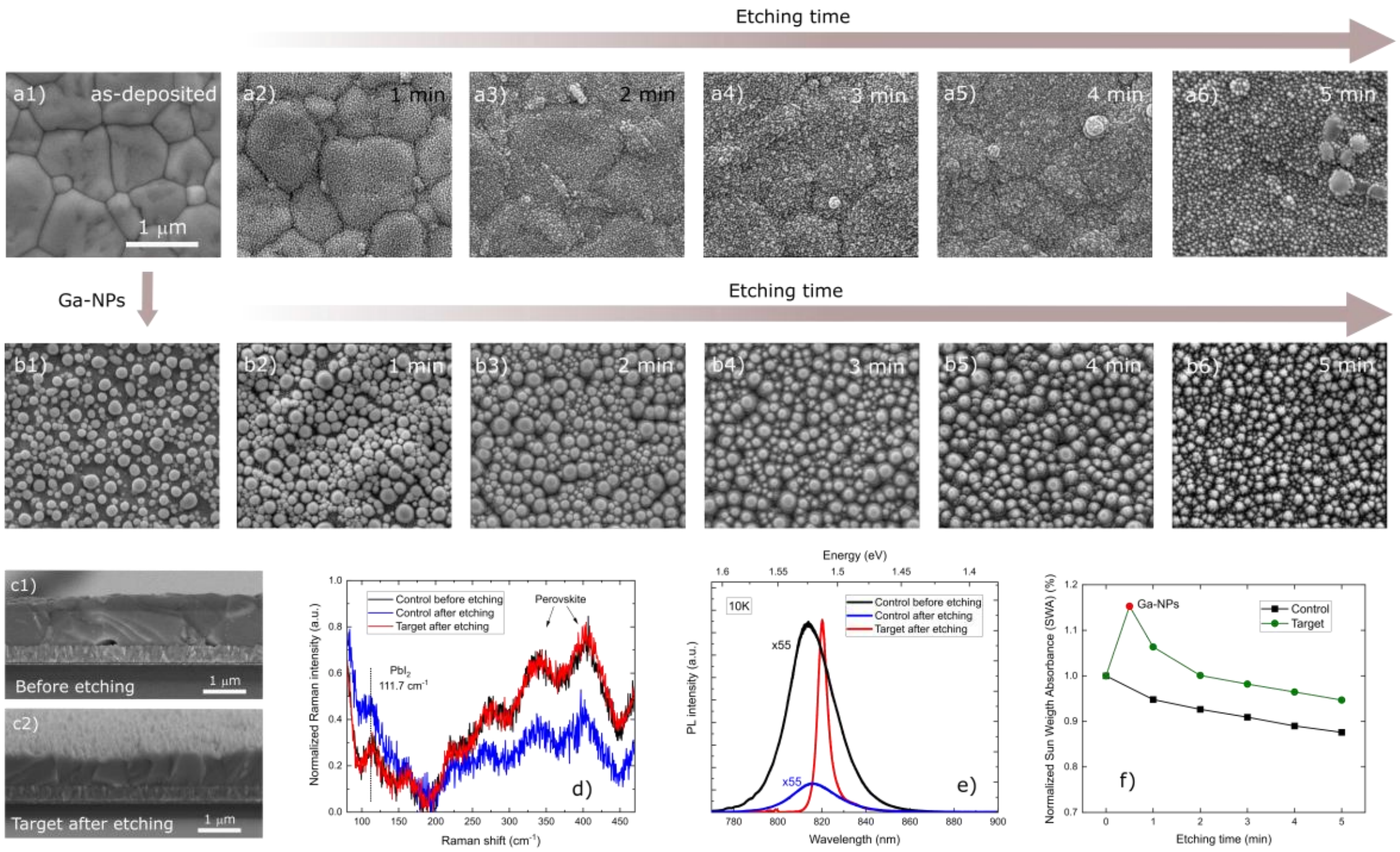

*Figure 4. Ga-NP-assisted top-down nanostructuring of polycrystalline $Cs_{0.05}FA_{0.85}MA_{0.1}PbI_3$ perovskite films. (a,b) Time-resolved top-view SEM comparison between bare perovskite control films and Ga-NP-coated target films subjected to identical Ar ICP-RIE treatments. (c) Cross-sectional SEM image of the nanostructured target film after complete Ga-mask consumption. (d) Raman spectra of the as-grown film, the etched control and the nanostructured target. (e) Low-temperature (10 K) photoluminescence spectra of the three samples. (f) Normalized sun-weighted absorbance as a function of etching time for control and target samples.*

The same protocol was also applied to thinner perovskite films prepared from 1.5 M precursors, with an initial thickness of approximately 530 nm, more representative of single-junction device absorbers. This study is included in Supplementary Note 3 and Figure S14. Although nanocone formation is retained, SWA decreases by 30-40% after only 3 min, highlighting the importance of initial film thickness for optical viability. After 5 min of etching, matching the treatment duration used for the 2 M films in Fig. 4, the contrast between control and target is visible to the naked eye (Figure S15 a): the etched control loses its characteristic dark appearance, whereas the Ga-NP-protected target remains dark. Consistently, X-ray diffraction (XRD) reveals that the perovskite peaks are no longer discernible in the etched control, while they remain clearly resolved in the target (Fig. S15b). Although the target exhibits lower diffraction intensities than the as-grown film, consistent with reduced film thickness, normalization to the perovskite (100) reflection reveals no appreciable increase in the relative $PbI_2$ peak intensity (Fig. S15 c). Similarly, no increase in the amorphous background is visible in the XRD pattern of the target, suggesting a negligible amorphization of the samples during the etching. These observations support the preservation of the crystalline perovskite phase during Ga-NP-assisted etching even in the thinner films, consistent with the Raman results for the 2M samples (Fig. 4 d), despite the substantial loss of optical absorption.

## Broadband optical response across semiconductors

Figure 5 summarizes the reflectance evolution during Ga-NP-assisted nanostructuring across the four semiconductor platforms. The optical measurements provide a proof of concept for light management using the self-consuming-mask approach, rather than an exhaustive optimization of antireflection performance for each material. The first column consists of the spectral reflectance while in the second column, the sun weight reflectance (SWR) according to equation 2 is included for each case where R(λ) is the reflectance spectra and S(λ) is the AM1.5G spectral photon flux.

$$SWR = \frac{\int S(\lambda) \cdot R(\lambda) \cdot d\lambda}{\int S(\lambda) \cdot d\lambda} \quad \text{Equation 2}$$

Starting with GaAs (Fig. 5 a)), whose polished surface delivers a baseline SWR of 26 ± 1%, characteristic of high refractive index of III-Vs. Depositing the self-assembled Ga-NPs of Fig. S5 (~115 ± 15 nm radius) yields a sharp rise to 43 ± 3%, attributable to metallic reflectivity compounded by LSPRs peaking in the near-IR range. Indeed, a minimum is observed around 400 nm corresponding to the out-of-plane LSPR mode where a maximum in located at longer wavelengths attributed to the in-plane plasmonic mode in agreement with previous works[18,28]. The error bar of the SWR reflects wafer-scale nonuniformity across a 2-inch quarter (Figure S16 a)). When Ar ICP-RIE initiates, there is a clear trend: between 0-300 s, the SWR drops up to 11% as hybrid Ga/GaAs nanostructures sharpen into nanocones. Nanocone pedestals rise amid inter-cone recessing, and Ga apices erode as demonstrated in previous sections. A mild increase is observed for longer times. A control experiment to a bare GaAs without Ga-NPs has been carried out for 300 s and included in Fig. S16 b) that confirms the benefits of nanostructures: the SWR decreases marginally, with rms even smoothing (2.2 to 1.7 nm in Fig. S16 c and d), underscoring physical nanocones sculpting over mere roughening.

GaP (Fig. 5 b)) exhibits a similar behavior. The bare substrate shows high SWR (~35%) due to the elevated refractive index. On the other hand, Ga-NPs increase it further via reflective LSPRs. However, etching procedures reduces reflectance broadband, with minima values around 12% over a 420-480 s region. For the case of the Si substrate (Fig. 5 c)), starting from ~38% SWR, Ga-NPs and post-etching improves the SWR to a ~15% minimum near 180 s.

In all three materials, the SWR values follow an identical pattern: a progressive decrease, minimum and then upturn. Critically, inspecting the nanostructures at each minimum reveals persistent Ga atop the cones, not pure semiconductor: for GaAs at 300 s (Fig. S5 b), hybrid Ga/GaAs with metallic caps; for GaP at 420 s (Fig. S6 a), analogous remnants; for Si at 180 s (Fig. S7 a), the same. As detailed in Supplementary Note 2, etching progressively reduces surface coverage, from initial >80% down through sparsening. To discern the role of this factor, Figure S17 replots SWR versus coverage (extracted from time-series corresponding SEM images), combining all three materials. Remarkably, despite contrasting etch times for each SWR minimum (180 s for Si vs 420 s for GaP), all these scenarios show almost identical coverage (~25-30%), irrespective of residual Ga volume. At these points, metallic-semiconductor hybrid nanostructures optimized index grading[1]. Persisting the etching beyond makes the coverage to decrease below ~20% (lower nanostructure density) that provokes low index-matching interaction and increases the SWR values.

The Reflectance light-management in perovskite material for the target sample is shown in Fig. 5 d). $CsFAMAPbI_3$ follows the same qualitative trend as a function of etching time but achieves dramatically lower reflectance than their monocrystalline III-Vs and Si counterparts. Specifically, Ga-NPs increases SWR of bare perovskite substrate from 21 to 28% via LSPR whose in-plane maximum is located at 640 nm in the VIS region. Then, when implementing the etching procedure, the SWR monotonically decreases, yielding a ~0.5% minimum upon complete Ga removal for 300 s of etching time. Spectral reflectance confirms uniform broadband suppression, with no parasitic features. The main point of perovskite nanocones to perform this ultralow SWR values lies in the high final coverage (82 ± 2%) quantified via 5×5 and 1×1 $\mu m^2$ AFM (Fig. S11 c) and d)), far exceeding the <30% of nanocone´s coverage in previous flat substrates. This dense perovskite arrays enables near-continuous index grading. Figure S18 details the light-management in the perovskite control for direct comparison. Reflectance spectra and SWR follow minor evolution. Bare film etching induces faint nanotexturing (rms from 22 to 27 nm in Fig. S11 a) and b)), dropping SWR to ~10% within the first 100 s before plateauing and no further reduction with time. Although the reduction of SWR in the control sample may seem worthy, it is important to remark the cost of severe absorbance degradation and the presence of $PbI_2$ (Fig. 4 d) and f)).

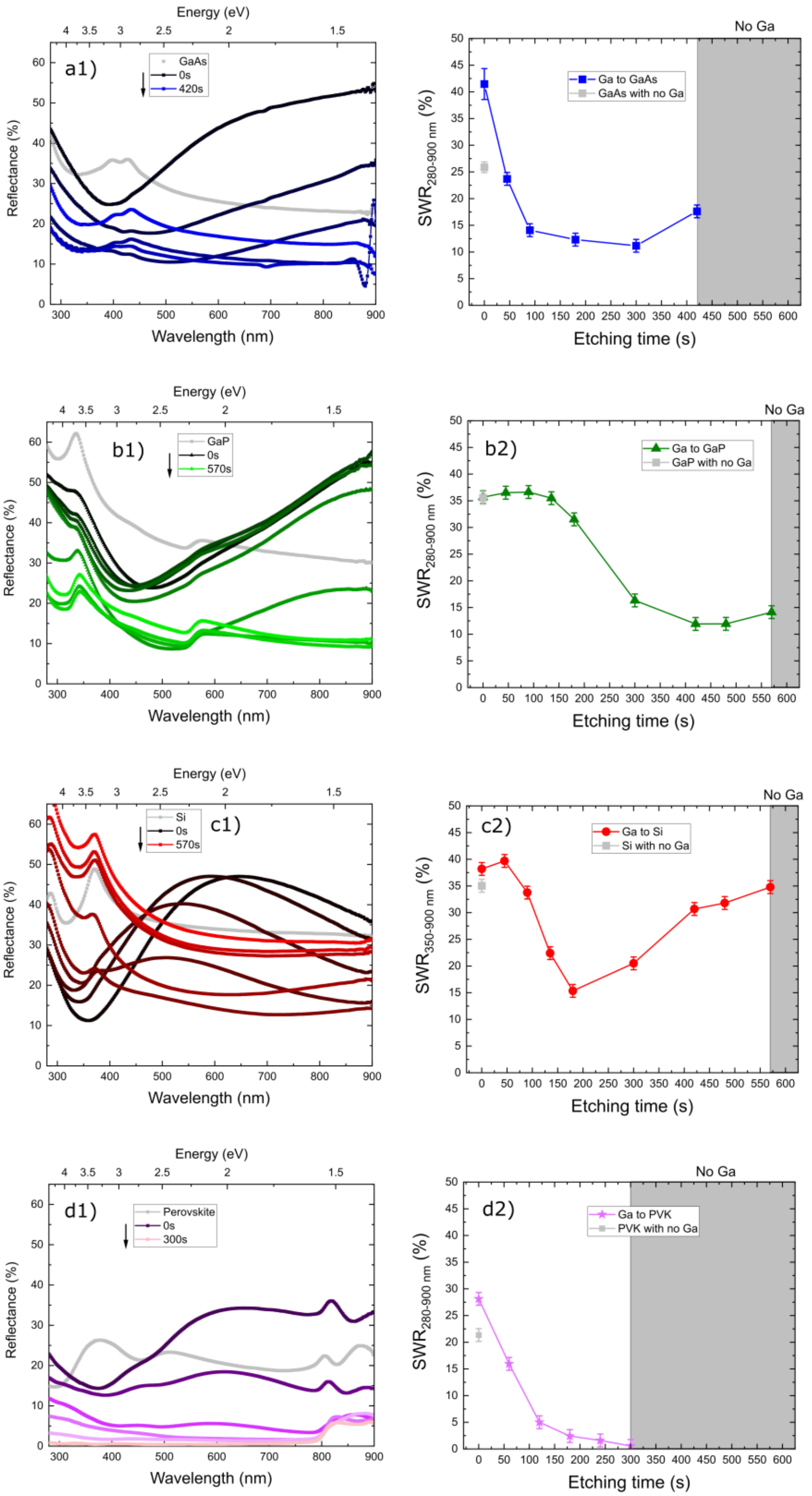

*Figure 5. Broadband reflectance suppression by Ga-NP-assisted nanostructuring across semiconductor platforms.(a–d) Spectral reflectance evolution, including the corresponding sun-weighted reflectance (SWR), during Ga-NP deposition and subsequent Ar ICP-RIE of (a) GaAs, (b) GaP, (c) Si and (d) $Cs_{0.05}FA_{0.85}MA_{0.1}PbI_3$ perovskite. Grey region in SWR spectra indicates when Ga is totally consumed.*

## Conclusions

Self-assembled Ga nanoparticles enable semiconductor nanocone formation through concurrent mask consumption and substrate etching during Ar ICP-RIE. Microscopy on GaAs resolves this mechanism and supports the preservation of crystalline continuity, while experiments on GaP, Si and polycrystalline perovskite demonstrate its transferability. Ga deposition and plasma conditions tune nanocone dimensions and coverage without a separate mask-removal step. The lowest sun-weighted reflectances measured under the investigated conditions were 11-15% on the monocrystalline substrates and 0.5% on the perovskite films. For the latter, comparison with directly etched controls supports the protective role of the transient mask, while experiments on thinner films highlight the remaining absorber thickness as a practical constraint. These results establish self-consuming Ga masks as a route to semiconductor light-management interfaces.

## Methods

**Semiconductor substrates preparation**

Monocrystalline substrates. GaAs (001) n+ substrates (Wafer Technology), Si (100) n+ substrates, and n-type GaP:S wafers (UniversityWafer) were sequentially cleaned by acetone, isopropanol and deionized water followed by $N_2$ blow-drying.

Perovskite thin film deposition. $Cs_{0.05}FA_{0.85}MA_{0.1}PbI_3$ perovskite films were fabricated from the powders precursors: CsI (19.5 mg), FAI (219.3 mg), MACl (11.2 mg), MAI (23.8 mg) and $PbI_2$ (722.8 mg). A solution of 1.5M and 2M were prepared by dissolving the precursors in DMF:DMSO (4:1 of mass ratio). The solution was stirred overnight at RT and filtered (PTFE, 0.22 µm) before deposition. The spin-coating was performed on NSG TEC15 FTO/glass substrates (12-14 $\Omega\cdot sq^{-1}$) with 400-450 nm thick and 12.5 nm of roughness. Prior to use, the FTO/glass substrate was sequentially cleaned by ultrasound devices in detergent, deionized water, acetone, and ethanol for 30 min each, followed by ultraviolet-ozone treatment for 30 min. Then, a drop of 60 µL of perovskite was deposited before ramping to 5000 rpm, 30s, ramp 1000 rpm/s). A drop of 200 µL of ethyl acetate as anti-solvent was placed at 25 s. Then, the samples were annealed at 100 °C for 30 min. All fabrication steps were carried out in a N2 glovebox (O2 and H2O < 0.01 ppm).

**Ga-NP deposition and ICP-RIE nanostructuring**

Ga-NPs were grown by Joule-effect thermal evaporation onto the different substrates. High-purity gallium (99.9999%) was evaporated in a vacuum chamber at a base pressure of $1\text{-}3\cdot10^{-6}$ mbar using a tungsten crucible operated at 50–60 W. The substrates were maintained at room temperature during deposition. The NP radius was controlled by adjusting the evaporation time using a shutter placed between the crucible and the substrate. The deposition rate was maintained at 1 $nm\cdot s^{-1}$, as monitored using a quartz crystal microbalance. Deposition times of 30 s were used for the TEM investigation of the Ga-NP-to-GaAs-nanocone transformation shown in Fig. 1 and for the coverage-evolution study. A deposition time of 150 s was used to produce the larger Ga-NPs for the size-scalability and associated AFM analyses on GaAs shown in Fig. 2, as well as for the experiments on GaP and Si. For the perovskite substrates, the deposition time was 80 s. After deposition, the samples were exposed to ambient air, forming a thin native oxide shell around the liquid Ga core.

The ICP-RIE procedure was performed with an Oxford Instruments PlasmaPro 100 equipment to create the nanostructures on the different substrates with and without Ga-NPs previously deposited. The plasma (Ar 30 sccm) was conducted by applying an ICP power of 100-400 W and a RIE power of 50 W in vacuum conditions (P=10 mTorr) being the sample holder at RT. The DC bias 500V during all processes. The etching procedure was controlled by time.

**Morphological and structural characterization**

Surface morphology of Ga-NPs on top of the different substrates was analyzed using a FEI Inspect F50 scanning electron microscope (SEM) operated at 5 keV. Size distribution histograms were extracted from SEM images using the ImageJ software. Top-view images were also used to assess existence of remaining Ga atop the hybrid nanostructures.

Atomic Force Microscopy (AFM) in tapping mode was used to obtain high-resolution topographic images and extract the RMS of the nanostructured surfaces using a Bruker Dimension Icon system. The commercial tapping mode Si cantilevers with a nominal tip radius of ≈8 nm, model ACTA-50, were purchased from AppNano, with a nominal resonance frequency in the range of 200-400 kHz and a nominal force constant in the range of 13-77 N/m. The Ga-NP images were analyzed with Gwyddion software to generate histograms of NP radii and to extract the Ga surface coverage over the substrate.

Cross-sectional TEM lamellae were prepared by FIB lift-out in a FEI FIB SCIOS dual-beam system after deposition of a protective Pt layer and in some cases by mechanical polishing followed by Ar+ ion milling in a Gatan PIPS. STEM imaging and EDX mapping were performed on a Thermo Fisher Talos F200X operated at 200 kV and equipped with a Super-X EDX system. Images were acquired in annular dark-field (ADF) mode, and FFT patterns were calculated from high-resolution STEM images. High-resolution ADF-STEM imaging of interfaces and nanocone facets was carried out on a Double-corrected FEI Titan Themis operated at 200 kV. Data was processed using Velox® and DigitalMicrograph® software using Velox® and DigitalMicrograph® software.

X-ray diffraction (XRD) profiles were recorded in Bragg-Brentano configuration using an Empyrean (Panalytical) diffractometer with a Cu Kα source, over a 2θ range of 10 to 70° and a step size of 0.02°.

**Optical characterization**

Reflectance spectra were recorded using a Jasco V-670 UV-Vis-NIR spectrophotometer equipped with an integrating sphere, at an incidence angle of 8° on 2 x 2 $cm^2$ samples. Measurements were performed over the 200-900 nm wavelength range, collecting both specular and diffuse reflectance. Absorbance spectra of the perovskite films were calculated from the measured transmittance as $A(\lambda) = -\log_{10}T(\lambda)$, with $T$expressed as a fraction. Sun-weighted reflectance (SWR) was calculated over 280-900 nm for perovskite, GaP and GaAs, and over 350-900 nm for Si, whereas sun-weighted absorbance (SWA) was calculated over 300-800 nm for the perovskite films. Both quantities were obtained by weighting the corresponding spectra with the AM1.5G solar photon flux and normalizing by the integrated photon flux over the respective wavelength interval. SWA values were subsequently normalized to that of the pristine, untreated perovskite control and expressed as percentages, with the reference assigned a value of 100%. These percentages represent relative changes in solar-weighted absorbance rather than absolute fractions of incident light absorbed. An apparent Urbach energy was estimated from the inverse slope of the approximately linear region of the logarithmic absorbance plotted against photon energy[27].

Raman spectroscopy was performed before and after etching using a HORIBA LabRAM HR spectrometer equipped with a 532 nm laser, a 100× microscope objective (NA = 0.90), an 1800 grooves·$mm^{-1}$ diffraction grating and a Si detector. The Raman shift was calibrated using a silicon reference substrate. Neutral-density filters with nominal transmissions of 10% and 0.1% were used for GaAs, GaP and Si, and for perovskite samples, respectively, yielding estimated incident laser powers of 3.5 and 0.035 mW at the sample. Each spectrum was acquired using an integration time of 30 s and three accumulations.

Photoluminescence (PL) measurements were performed at 10 K using a separate PL module coupled to the same HORIBA LabRAM HR spectrometer. Excitation was provided by a 532 nm laser through a 50× microscope objective (NA = 0.90), using a neutral-density filter with a nominal transmission of 0.1%, yielding an estimated incident laser power of 0.0157 mW at the sample. Each spectrum was acquired using an integration time of 30 s and three accumulations.

## Competing Interests Statements

S.C.G., A.H.C., J.U.H., V.B.B. and D.G.R. are named inventors on granted Spanish patent ES3009234 B2, entitled "Procedimiento para producir nanopartículas semiconductoras tridimensionales a partir de nanopartículas metálicas autoensambladas", held by Universidad Politécnica de Madrid and Universidad de Cádiz and relating to the nanostructuring method reported in this work. The remaining authors declare no competing interests.

## Acknowledgements

This work was supported by the Madrid Government (Comunidad de Madrid, Spain) through the Multiannual Agreement 2023-2026 with Universidad Politécnica de Madrid, under Line A, Emerging PhD Researchers, through the CADENCE project, by project FEDER-UCA-2024-A2-08, funded by the Programa Operativo FEDER

Andalucía 2021-2027 and the Consejería de Universidad, Investigación e Innovación, Junta de Andalucía. The work was also supported by Universidad Rey Juan Carlos "Proyecto Puente" project with reference F1403. The authors acknowledge the instrumentation and technical advice provided by the National Facility ELECMI ICTS, through its División de Microscopía Electrónica (DME) node at Universidad de Cádiz, under access reference ELC667-2025 from the ELECMI 2024 call. The authors also acknowledge the facilities and technical support provided by the ISOM node of ICTS Micronanofabs at Universidad Politécnica de Madrid, as well as the Center of Technological Support (CAT) at Universidad Rey Juan Carlos.

## Data availability

The data supporting the findings of this study are available from the corresponding author upon reasonable request.

# Supporting Information of Nanostructuring Semiconductor Surfaces Using Self-Assembled Metal Nanoparticles for Optoelectronic Applications

S. Catalán-Gomez[1,2]*, M. Ibáñez[1], P. Guo[3], V. Braza[4], B. Vázquez[1], R. Sun[3], F. Li[3], P. Ibáñez[1], Y.K. Ryu[1], M. Vila[5], J. Bartolomé[5], D.F. Reyes[4], J. Martínez[1], D. González[4], H. Wang[3], J.M. Ulloa[1] and A. Hierro[1]

[1]Instituto de Sistemas Optoelectrónicos y Microtecnología, Universidad Politécnica de Madrid, ETSI Telecomunicación, Av. Complutense 30, E-28040 Madrid, Spain

[2]Departamento de Ingeniería Eléctrica, Electrónica Automática y Física Aplicada, Universidad Politécnica de Madrid, ETSI y Diseño Industrial, Ronda de Valencia 3, E-28012 Madrid, Spain

[3]State Key Laboratory of Solidification Processing, Center for Nano Energy Materials, Northwestern Polytechnical University, Xi'an, P. R. China.

[4]Departamento de Ciencia de los Materiales e IM y QI, Universidad de Cádiz, E-11510 Puerto Real, Cádiz, Spain

[5]Escuela de Ingeniería de Fuenlabrada, Universidad Rey Juan Carlos, Fuenlabrada, E-28942 Madrid, Spain

*Corresponding author:

sergio.catalan.gomez@upm.es

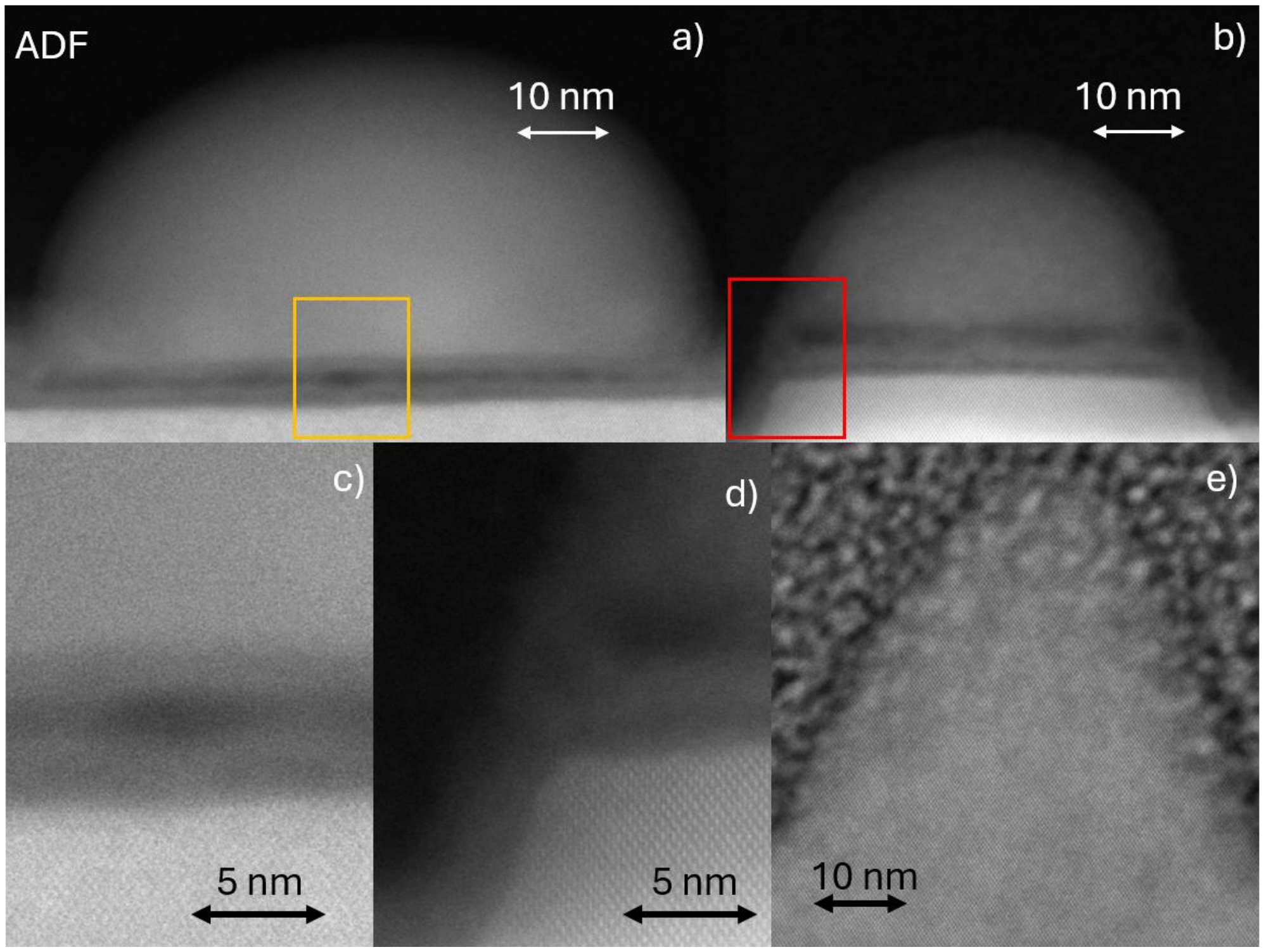

*Figure S1. High-resolution ADF-STEM analysis of surface amorphization. (a) Ga-NP on GaAs before etching, showing the native amorphous interfacial layer between droplet and substrate. (b) Hybrid Ga/GaAs nanostructure at an intermediate etching stage (Ar plasma). (c,d) Magnified views of the regions marked in (a) and (b). The amorphous layer measures approximately [4.9 ± 0.5] nm beneath the unetched Ga-NP, [5.4 ± 0.4] nm beneath the partially consumed Ga-NP and [3.6 ± 0.3] nm on the exposed pedestal surface. (e) Fully formed GaAs nanocone after Ar ICP-RIE, showing lattice fringes extending up to the side facets.*

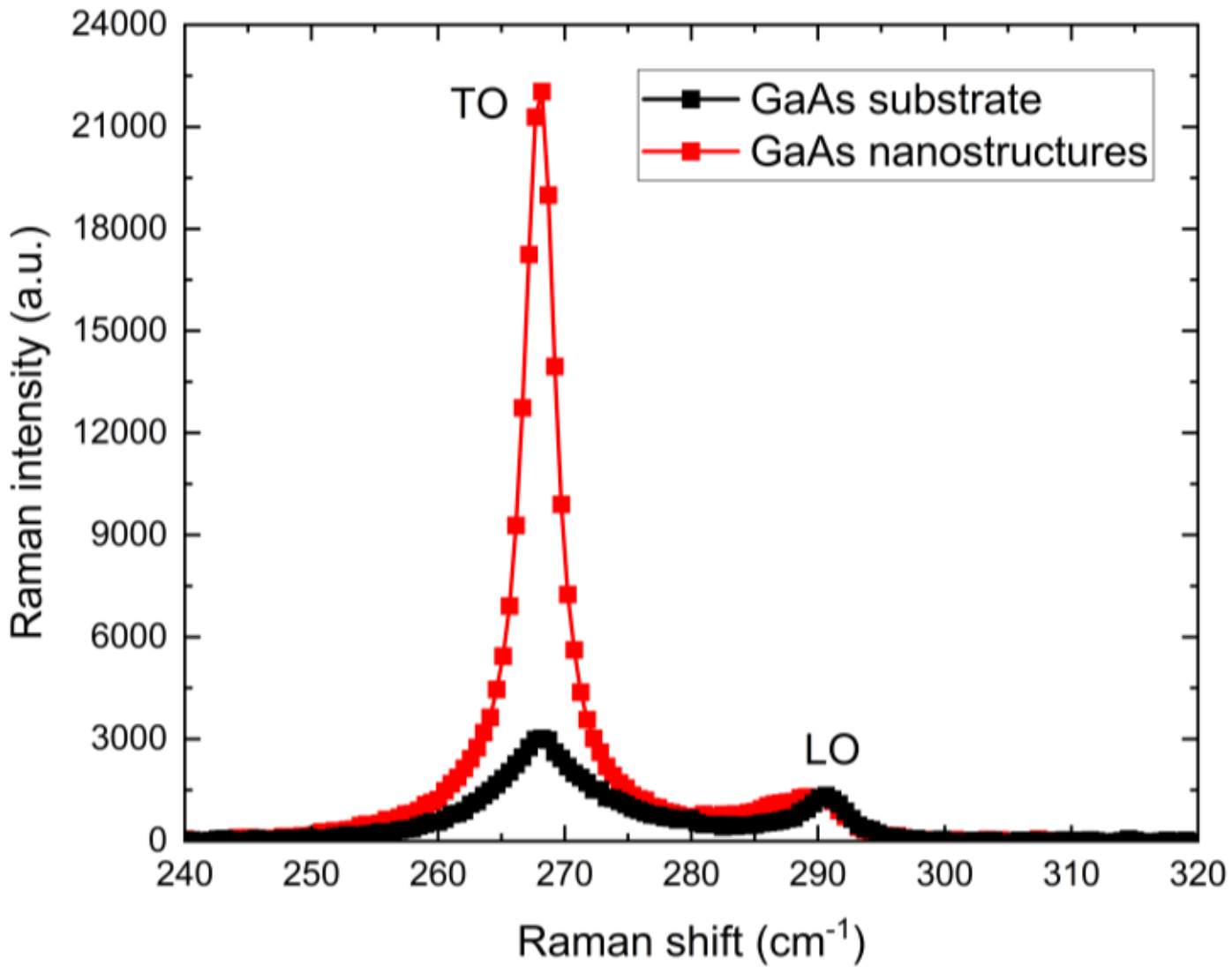

*Figure S2. Raman spectra of flat GaAs (black squares) and GaAs nanostructures of Fig. 1 d) (red squares) measured under identical conditions.*

## Supplementary Note 1: Calibration of GaAs and Ga etching rates

The self-consuming mechanism relies on etching rates of the same order of magnitude between Ga-NPs and the underlying GaAs substrate under Ar ICP-RIE conditions. Here we quantify these rates independently and demonstrate their similarity.

GaAs etch rate: We patterned GaAs with AZ5214 photolithographic resist under standard conditions. Then, we etched the whole surface under the standard ICP-RIE conditions for 4 different times (50, 100, 150 and 200s). Finally, we make the removal of the resist with acetone and measured the resulting step height by AFM. A sketch of the calibration experiment is shown in Figure S3 a) while the AFM results are included in b). Linear regression (Fig. S3 c) of step depth versus etch time yields an etch rate of 0.69 ± 0.03 nm $s^{-1}$ (n = 5, $R^2$ = 0.99).

Ga etch rate: From TEM results of the Fig. 1 time series, we measured Ga-NP apical radii at intermediate stages. After 50 s, remnant Ga-NPs have radii of 22 ± 2 nm while after 100 s only NPs of 8 ± 3 nm (Fig. S3c). Linear fitting gives a Ga radial etch rate of 0.43 ± 0.01 nm $s^{-1}$. Assuming a isotropic erosion, an effective horizontal rate of ≈0.8 nm $s^{-1}$ would be obtained accounting for geometry.

The GaAs rate exceeds Ga by a factor of ≈1.6 (selectivity), sufficient for nanostructuring and enabling timely mask consumption.

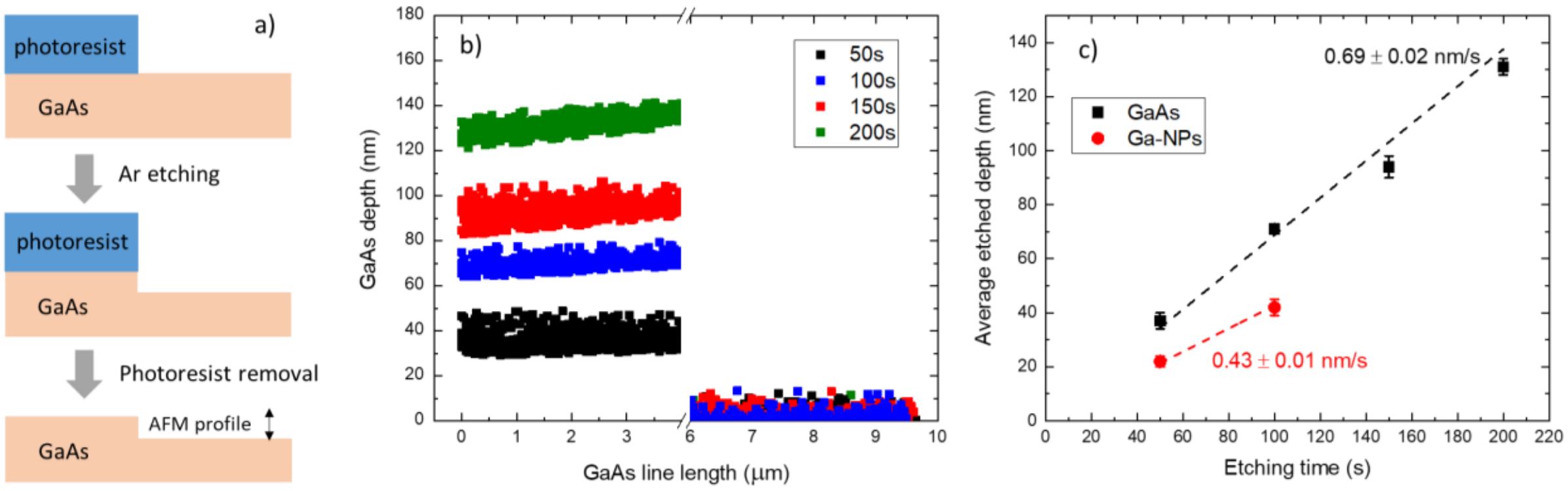

*Figure S3. a) Sketch of the GaAs etching rate calibration experiment. b) GaAs profile after calibration experiment extracted by AFM measurements for different etching times. c) Average etched depth versus etching time for GaAs and Ga-NP radius decrease from TEM analysis of Figure 1. Linear fitting of the data is included for each case.*

## Supplementary Note 2: Coverage as a function of etching time

The AFM images in Fig. 2 a) qualitatively illustrate the progressive sparsening of the Ga-NP array into GaAs nanocones. Here we quantify this evolution using a complete time series of samples prepared similarly to those in Fig. 1 (small Ga-NPs).

Figure S4 a) presents 5×5 $\mu m^2$ AFM images of Ga-NPs etched at 0, 30, 90, 150 and 210 s. Before etching (t=0 s), the as-deposited sample appears as a dense array with Ga-NPs coating representing 85% of the surface coverage. By 30 s, recesses emerge between NPs as GaAs etches vertically, while NPs also shrink laterally and coverage drops to ~80%. At 90 s, hybrid nanostructures dominate (Ga caps atop GaAs bases) with coverage ~59%. By 150 s (Fig. S4 a4), Ga is fully consumed (no metallic presence in TEM Fig. 1 d), leaving a sparse nanocone array (coverage ≈28%). Extended etching to 210 s further reduces coverage to ~5%, as GaAs nanocones are still sputtered. The quantification of the coverage as a function of etching time is shown in Fig. S4 c) with different regions marked within the graph corresponding to the material´s constituent of the nanostructure.

The evolution of NP size distributions through their histograms provides mechanistic insight into the coverage decay. Upon deposition (t=0 s), the NP base radius histogram (Fig. S4 b1)) exhibits a characteristic bimodal distribution driven by Ostwald ripening and coalescence during Ga dewetting, where larger NPs grow at the expense of smaller ones. In this case, (30 s of evaporation time), it yields to stable large droplets (mean radius ~37 nm) surrounded by smaller NPs (mean ~9 nm). As etching progresses (30 an 90s), the small-NP population vanishes rapidly, while large NPs still resist. By 150 s (Fig. S4 b4)), the distribution ends to GaAs nanocones with a unimodal form peaked at ~24 nm. Critically, the NPs density (y-axis counts/$\mu m^2$) drops from 15 NPs/$\mu m^2$ initially to ~10 NPs/$\mu m^2$ at 150 s of etching time, reflecting both shrinkage and inter-NP recessing (lower coverage).

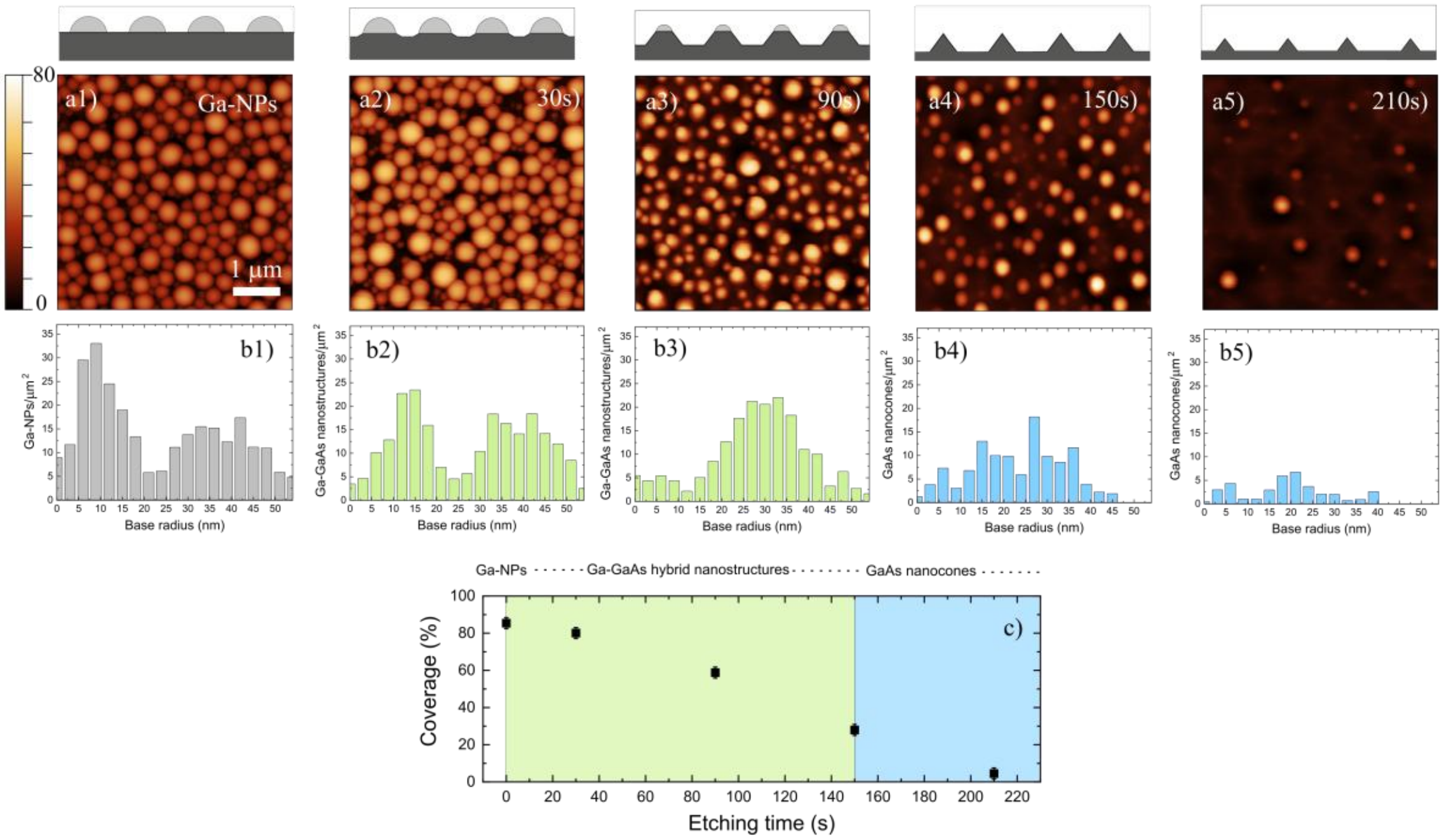

*Figure S4. Evolution of surface morphology, nanostructure size distribution and surface coverage during the conversion of small Ga-NPs into GaAs nanocones. (a) AFM images acquired after 0, 30, 90, 150 and 210 s of Ar ICP-RIE. (b) Corresponding histograms of the nanostructure base radius. (c) Surface coverage as a function of etching time, identifying the initial Ga-NP, hybrid Ga/GaAs and fully converted GaAs-nanocone regimes. The progressive loss of the small-NP population and the lateral shrinkage of the remaining structures reduce the coverage from approximately 85% to 5% for a total of 210 s of etching time.*

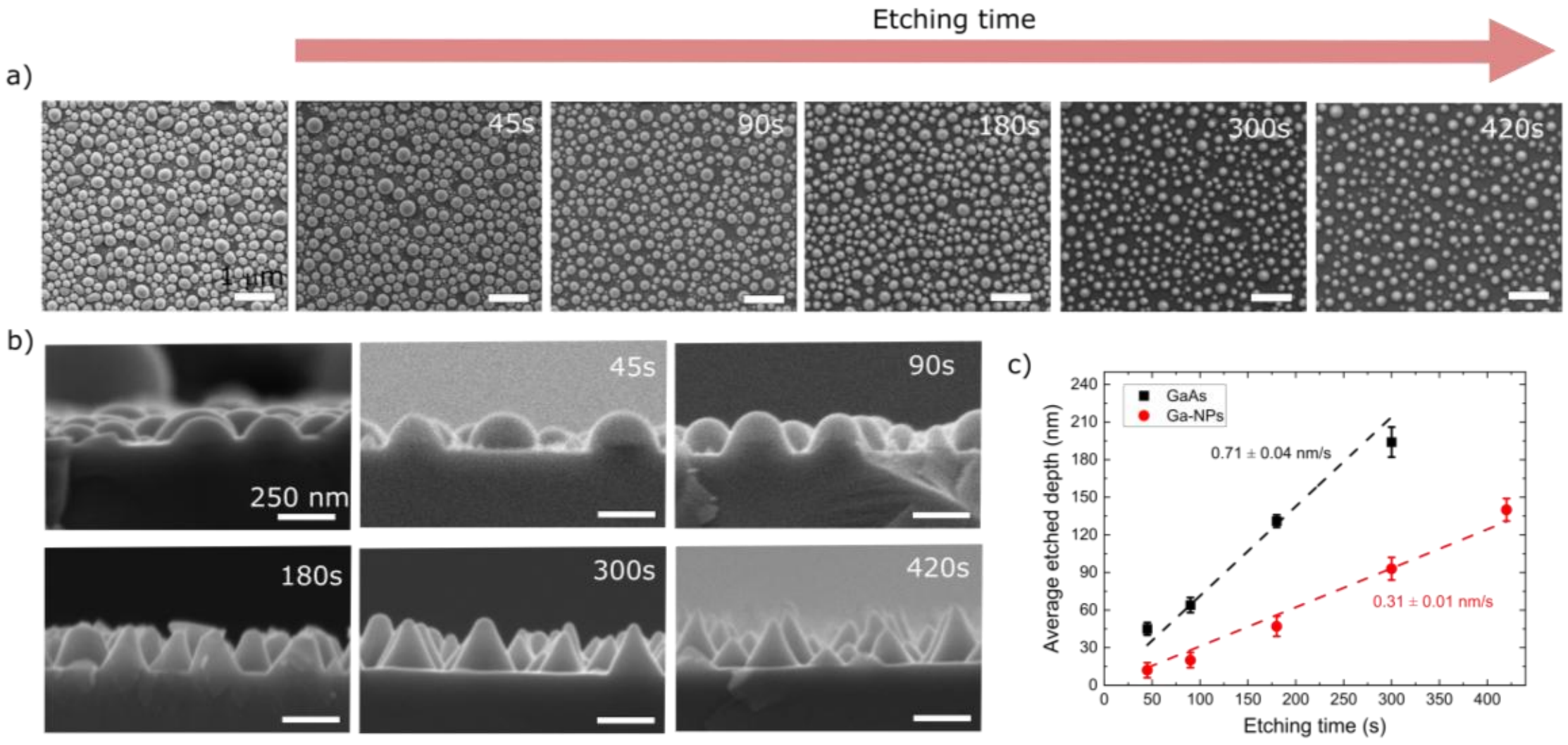

*Figure S5. Time-resolved morphological evolution of large Ga-NPs on GaAs during Ar ICP-RIE. (a, b) Top-view and cross-sectional SEM images acquired at successive etching times, showing the transition from Ga droplets to hybrid Ga/GaAs*

*structures and, finally, GaAs nanocones. (c) Evolution of Ga-NP and GaAs pedestal dimensions with etching time extracted from cross-sectional SEM analysis. Dashed lineal fitting indicates the etching rate of each material at this specific conditions. Scale bars in a) correspond to 1 μm and correspond to 250 nm in b).*

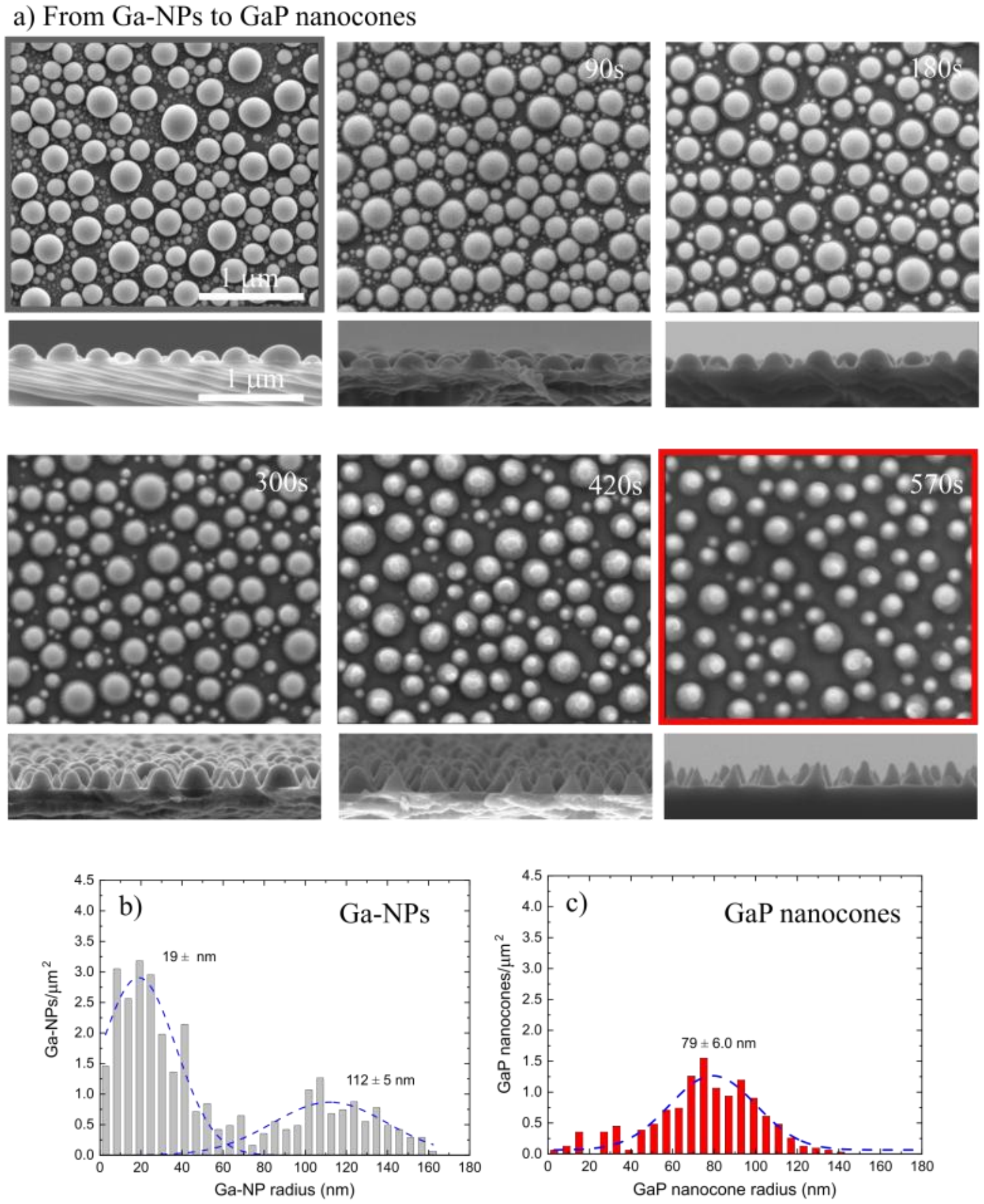

*Figure S6. Morphological evolution of large Ga-NPs on GaP during Ar ICP-RIE. (a) Top-view and cross-sectional SEM images acquired at successive etching times. (b) Size distribution of the as-deposited Ga-NPs on GaP. (c) Size distribution of the resulting GaP nanocones after complete mask consumption for 570s.*

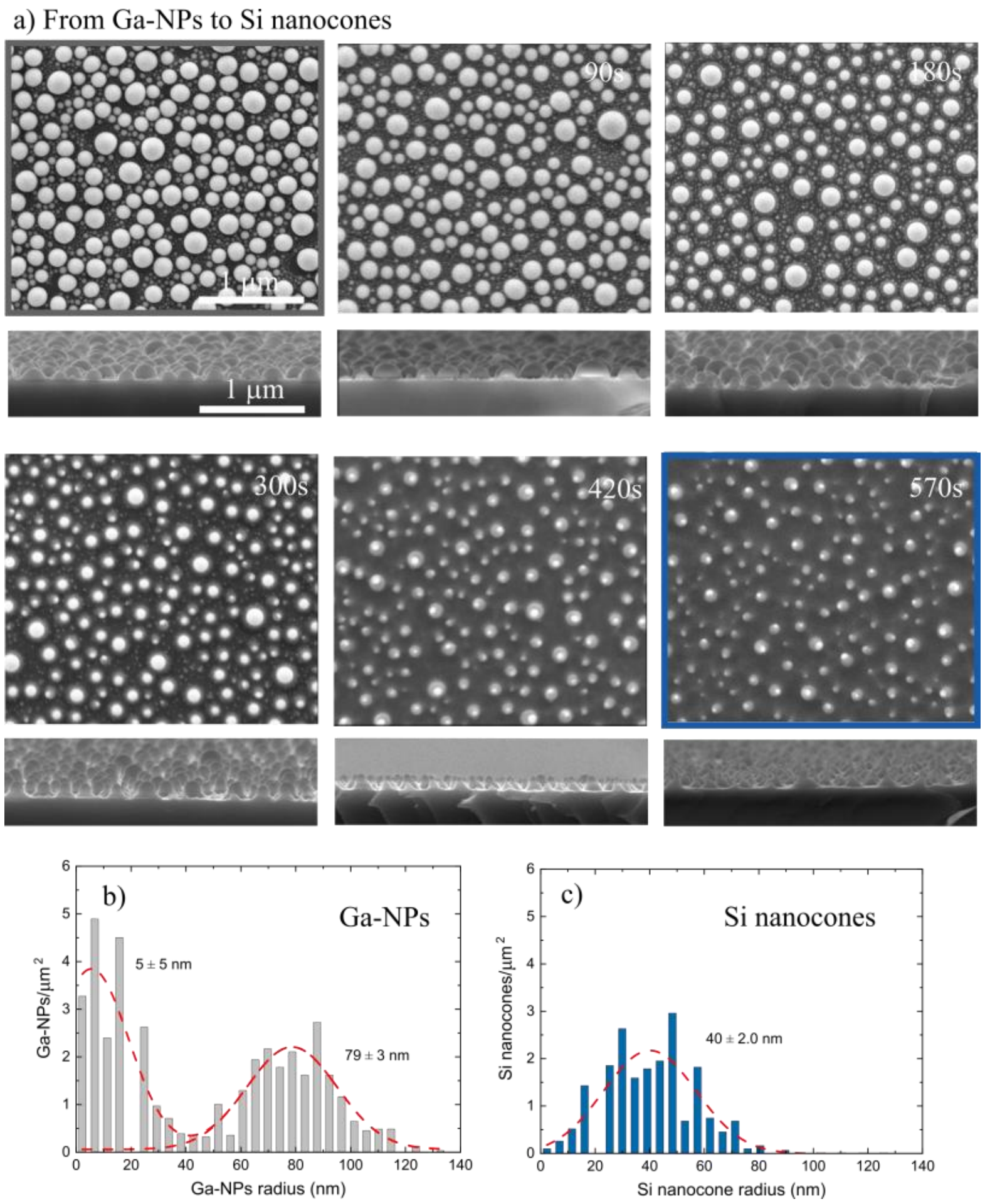

*Figure S7. Morphological evolution of large Ga-NPs on Si during Ar ICP-RIE. (a) Top-view and cross-sectional SEM images acquired at successive etching times. (b) Size distribution of the as-deposited Ga-NPs on Si. (c) Size distribution of the resulting Si nanocones after complete mask consumption for 570s.*

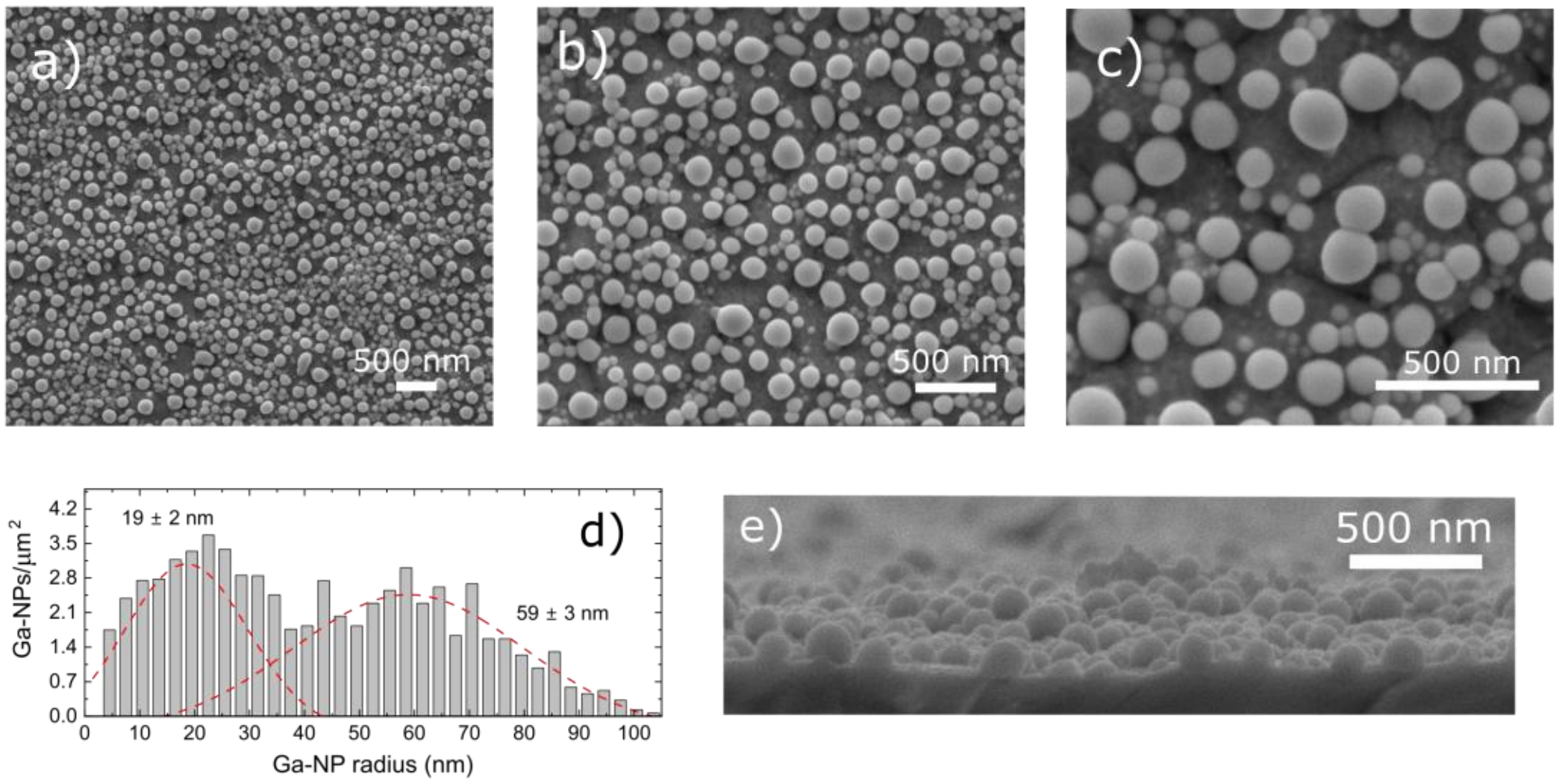

*Figure S8. Morphological characterization of the Ga-NP-coated $CsFAMAPbI_3$ target before Ar ICP-RIE. (a–c) Top-view SEM images at increasing magnification of the as-deposited Ga-NP mask on the perovskite surface. (d) Corresponding Ga-*

*NP size distribution. (e) Cross-sectional SEM image showing the Ga-NP morphology and contact with the polycrystalline perovskite film. Scale bars are indicated in the panels.*

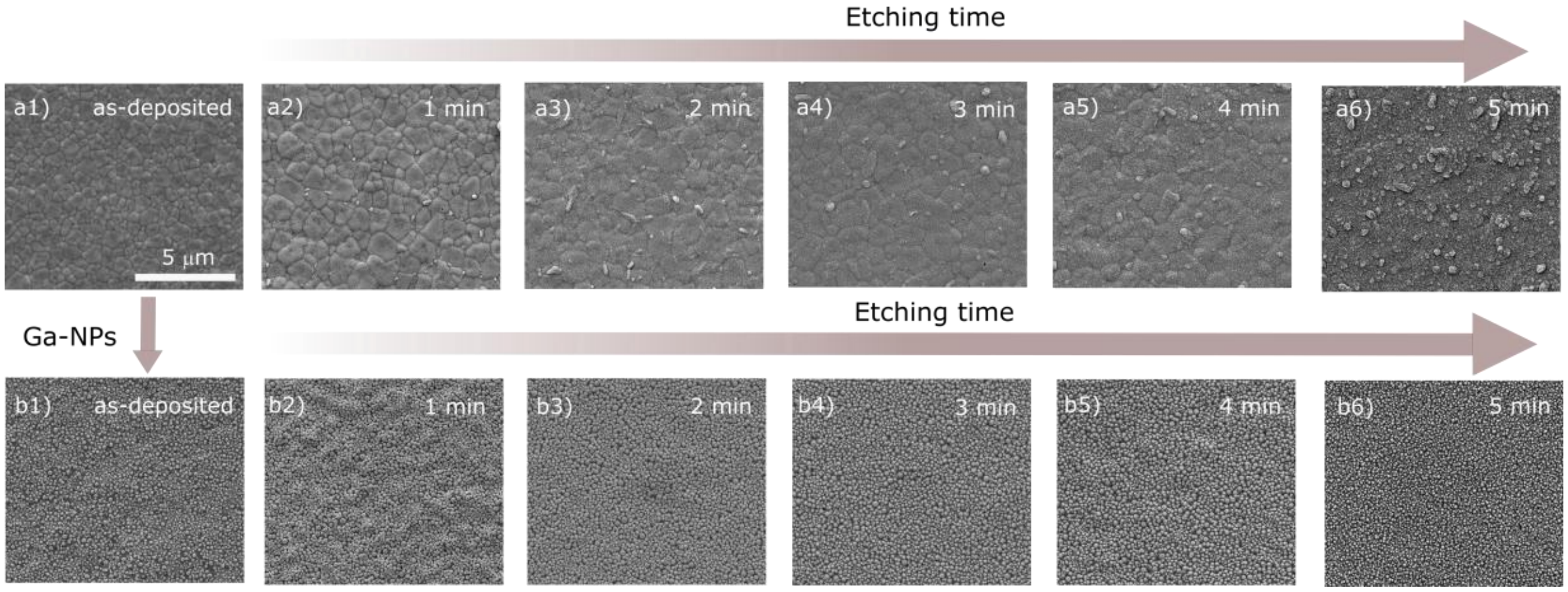

*Figure S9. Low-magnification top-view SEM images of (a) bare perovskite control films and (b) Ga-NP-coated target films subjected to increasing Ar ICP-RIE times. The control evolves into a weakly roughened and irregular surface, whereas the target develops a homogeneous high-density nanocone array. All images have the same size with the scale bar corresponding to 5 µm.*

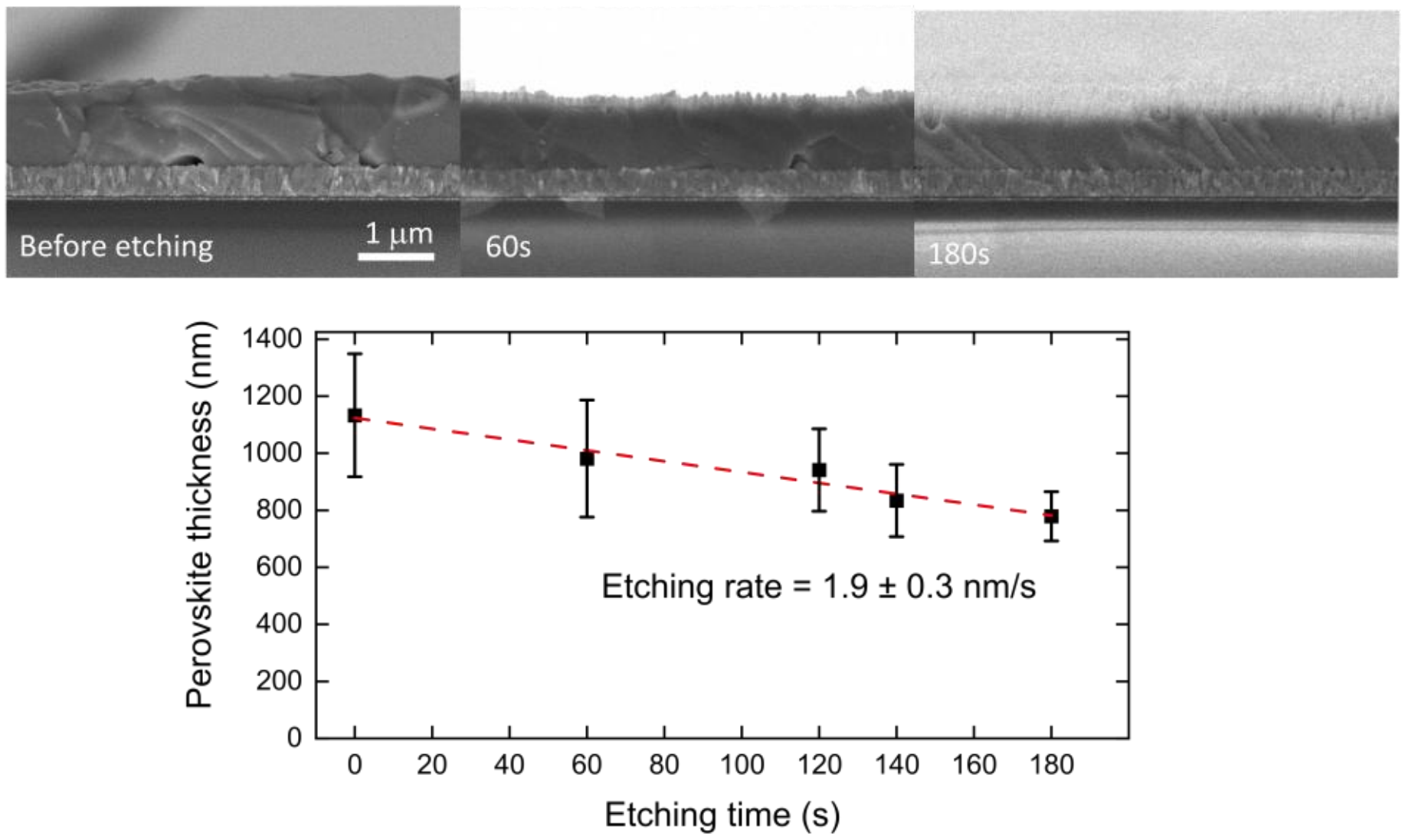

*Figure S10. Cross-sectional SEM calibration of the $CsFAMAPbI_3$ perovskite etching rate under the standard Ar ICP-RIE conditions. Representative images before etching and after 60 and 180 s are shown at the top. The lower panel summarizes thickness measurements acquired at more than 50 locations across 1.5×1.5 cm samples for etching times of 0, 60, 120, 140 and 180 s. Linear regression yields an etching rate of 1.9±0.3 $nm{\cdot}s^{-1}$. Error bars represent the standard deviation arising primarily from the initial thickness non-uniformity of spin-coated films.*

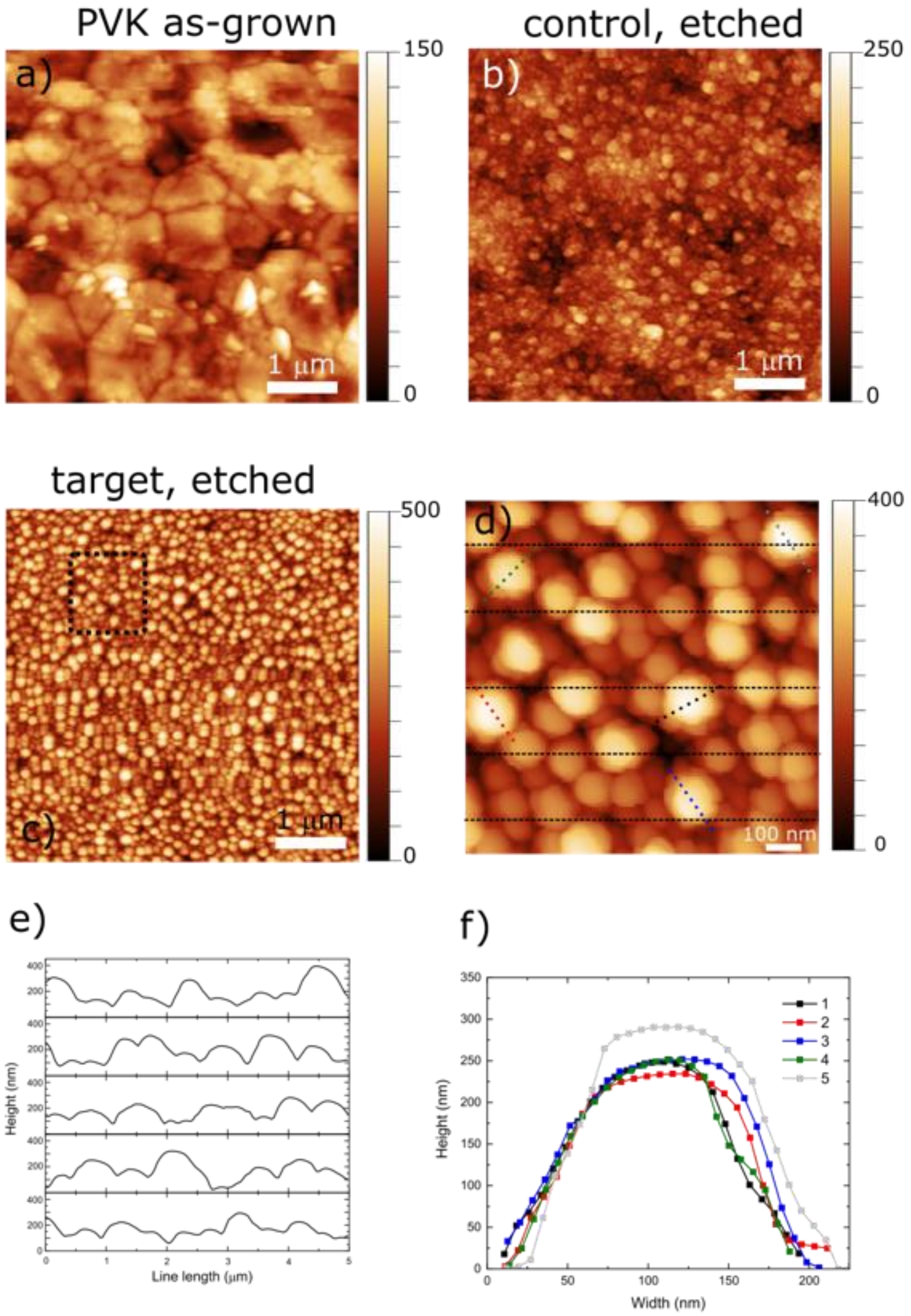

*Figure S11. AFM analysis of the perovskite surface morphology before and after Ar ICP-RIE. (a–c) Large-area AFM height images of the as-grown perovskite film, etched control and Ga-NP-assisted nanostructured target, respectively. d) Higher-resolution height image, e) representative line profiles of the corresponding horizontal lines in d) and f) marked nanostructures in d). The target sample displays nanocones with typical base diameters of 100-150 nm and an aspect ratio of approximately 1.25, whereas direct etching of the control only produces minor roughening.*

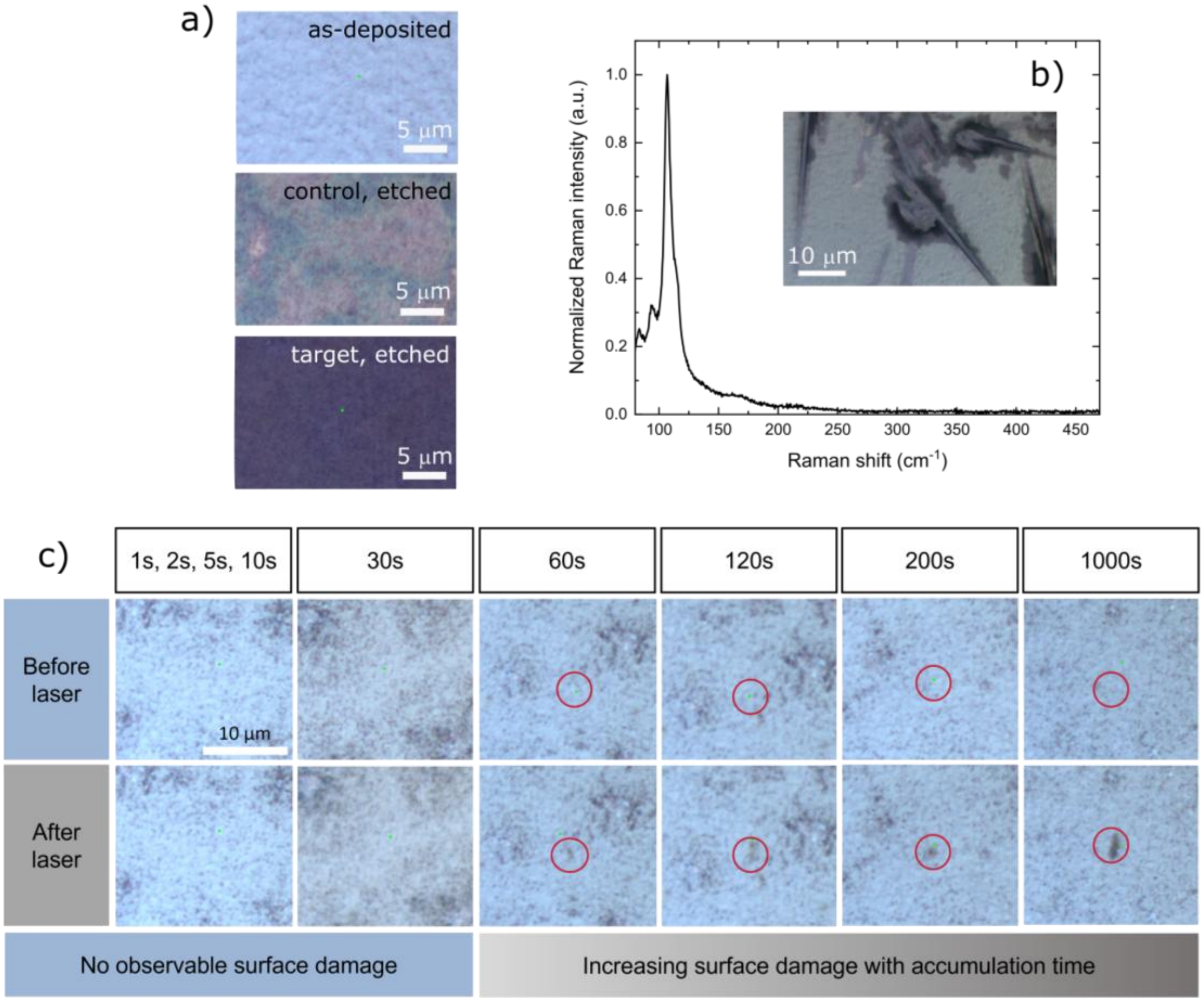

*Figure S12. Raman-control measurements for the perovskite samples. (a) Optical microscope images of the as-grown control, etched control and nanostructured target acquired under the Raman measurement conditions. (b) Raman spectrum of a $PbI_2$ reference crystal, showing the characteristic phonon mode near 111.7 $cm^{-1}$. (c) Evolution of the Raman microscope image of the as-grown perovskite as a function of Raman measurement acquisition time, used to identify conditions that avoid laser-induced degradation.*

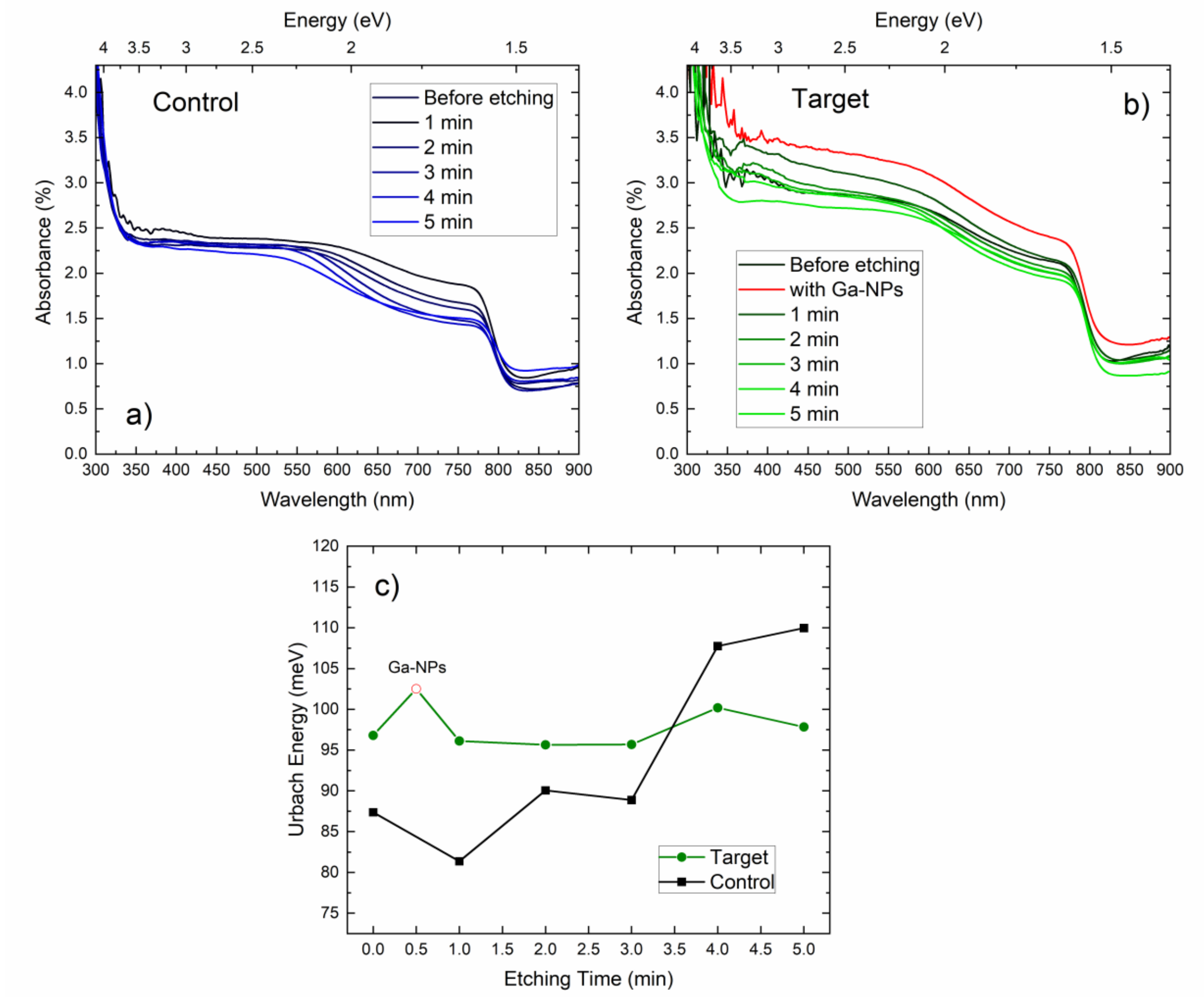

*Figure S13. (a,b) Absorbance spectra of the control and Ga-NP-coated target films, respectively, acquired at successive Ar RIE-ICP etching times. (c) Urbach energy extracted from the absorption-edge tail as a function of etching time. The target retains a nearly constant Urbach energy, whereas the direct-etched control shows a progressive increase after prolonged plasma exposure.*

## Supplementary Note 3: Nanotexturing of 1.5M perovskite.

The results presented in Figure 4 for the ~1 µm thick $CsFAMAPbI_3$ film (2 M precursor) demonstrate robust nanocone formation with no detectable degradation. Here we investigate whether the same Ga-NP protocol applies to thinner films more representative of high-efficiency single junction perovskite solar cells.

We fabricated control and target samples from 1.5 M precursors, yielding initial thicknesses of ~530 nm with grain sizes of 350 nm as observed in Figure S14 a1). Then, as target sample, Ga-NPs were deposited under the same conditions as Figure 4. The surface evolution as a function of etching in the SEM images of first and second row reveals that nanocone formation proceeds morphologically identically to the thicker films: the target sample develops nanocones after 3 min etch (Ga mask not fully consumed), while the control shows only faint roughening transitioning to $PbI_2$-like features (Fig. S14 a 4)).

Absorbance spectra (Fig. S14 c) and d)) and normalized SWA values for each case (Fig. S14 e)) highlight the vulnerability of thin films. After only 3 min, SWA drops by ~30% (target) and ~40% (control), far exceeding the 6 and 13% relative losses in 1 µm films (Fig. 4 f)). Spectral shapes confirm $PbI_2$ emergence in control samples (persistent 300-500 nm tail) but uniform decreasing in the target case.

Furthermore, Urbach energy (Fig. S14 f)) rises more aggressively than in thick films, particularly in control (>30% increase), indicating bulk disorder. This aligns with SEM ($PbI_2$ structures) and indirect spectral evidence, underscoring that while morphology transfers, optical viability demands sufficient initial thickness to buffer plasma-induced losses.

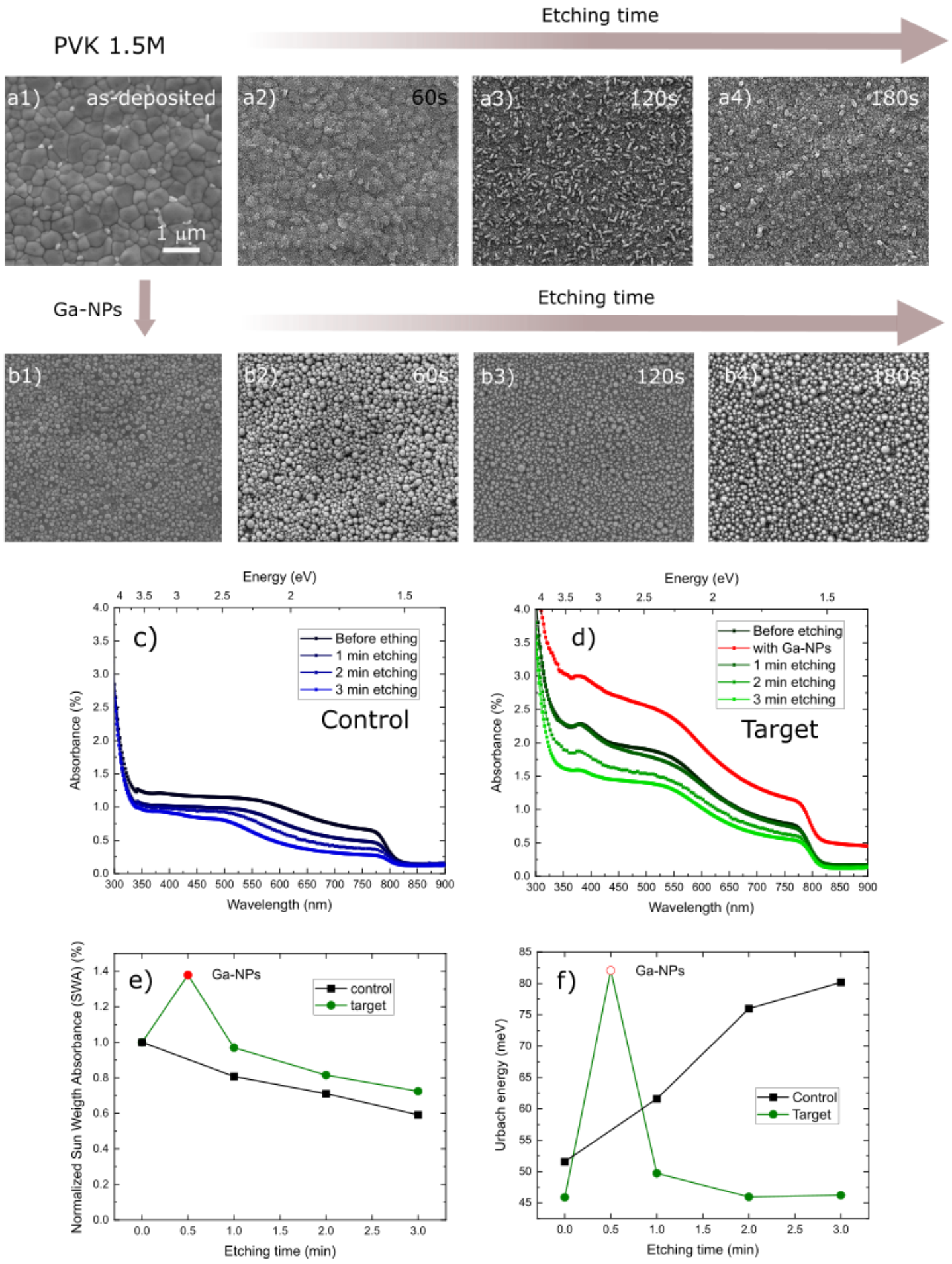

*Figure S14. Nanostructuring of thinner perovskite films prepared from 1.5 M precursor solution. (a,b) Top-view SEM images of bare control and Ga-NP-coated target films, respectively, before and after increasing Ar ICP-RIE times. The target develops nanocones after 180 s, while direct etching of the control yields irregular degradation features. (c,d) Absorbance spectra of the control and target samples as a function of etching time. (e) Normalized sun-weighted absorbance and (f) Urbach energy evolution for both sample types. Although the morphology is transferred to the thinner target film,*

*the rapid absorbance loss highlights the importance of sufficient initial perovskite thickness. Scale bars are indicated in the panels.*

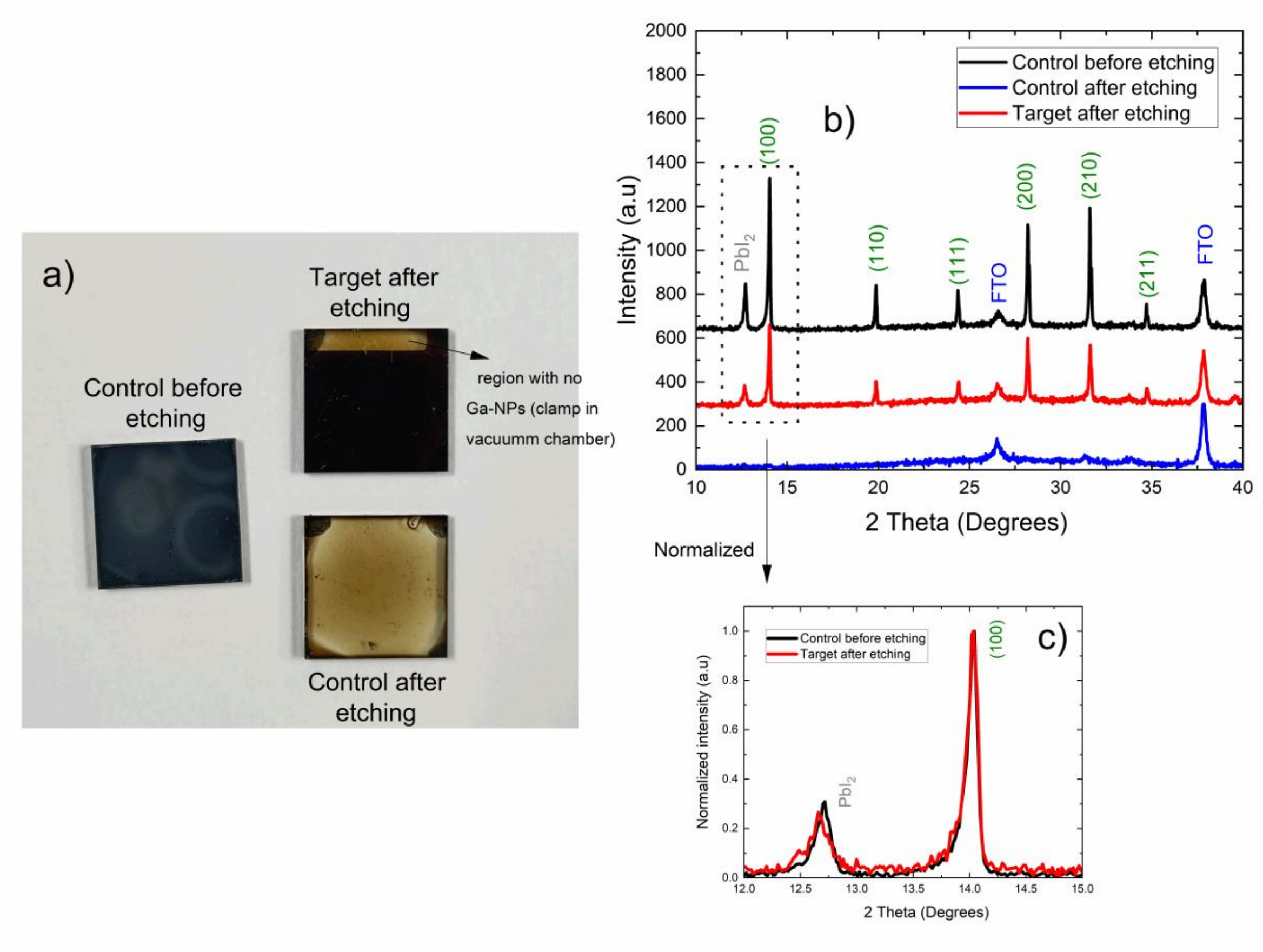

*Figure S15. (a) Photograph of perovskite films prepared from 1.5 M precursors, showing the as-grown control and the control and target samples after 5 min of Ar ICP-RIE. The region shielded by the sample holder during Ga-NP deposition is indicated on the target. (b) XRD patterns of the three samples, with perovskite, PbI2 and fluorine-doped tin oxide (FTO) substrate XRD peaks identified. The dashed box marks the angular range zoom in (c). (c) Comparison of the as-grown control and etched target patterns normalized to the perovskite (100) peak, showing no appreciable increase in the relative PbI2 reflection after Ga-NP-assisted etching.*

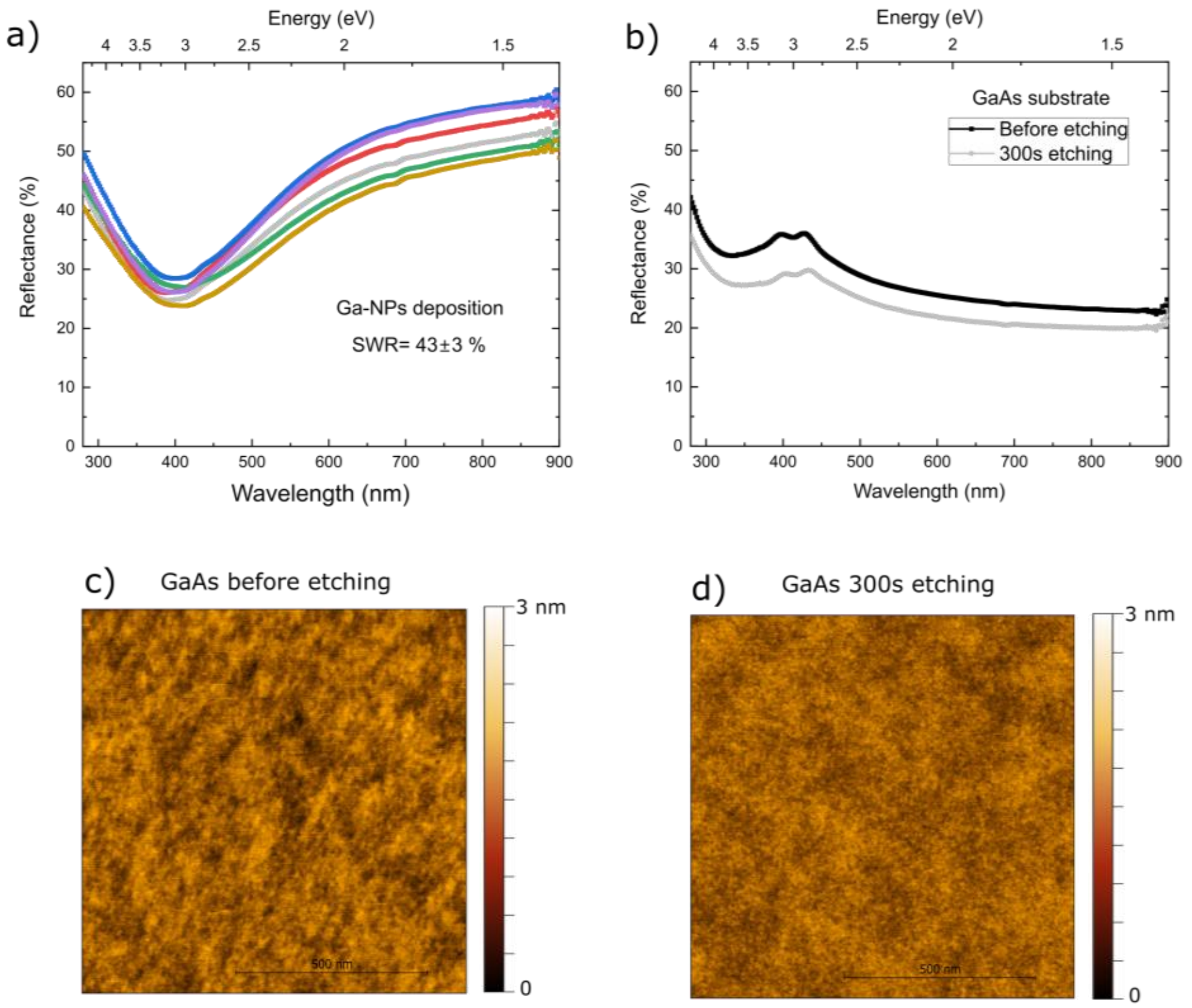

*Figure S16. Spatial uniformity and control experiment for GaAs reflectance reduction. (a) Reflectance measured at multiple positions across a quarter of a 2-inch GaAs wafer after Ga-NPs deposition. (b) Reflectance spectra of flat GaAs before and after 300 s of direct Ar ICP-RIE without Ga-NPs. (c,d) AFM height images of the corresponding uncoated GaAs surfaces, showing that direct plasma exposure slightly smooths the surface rather than forming nanocones.*

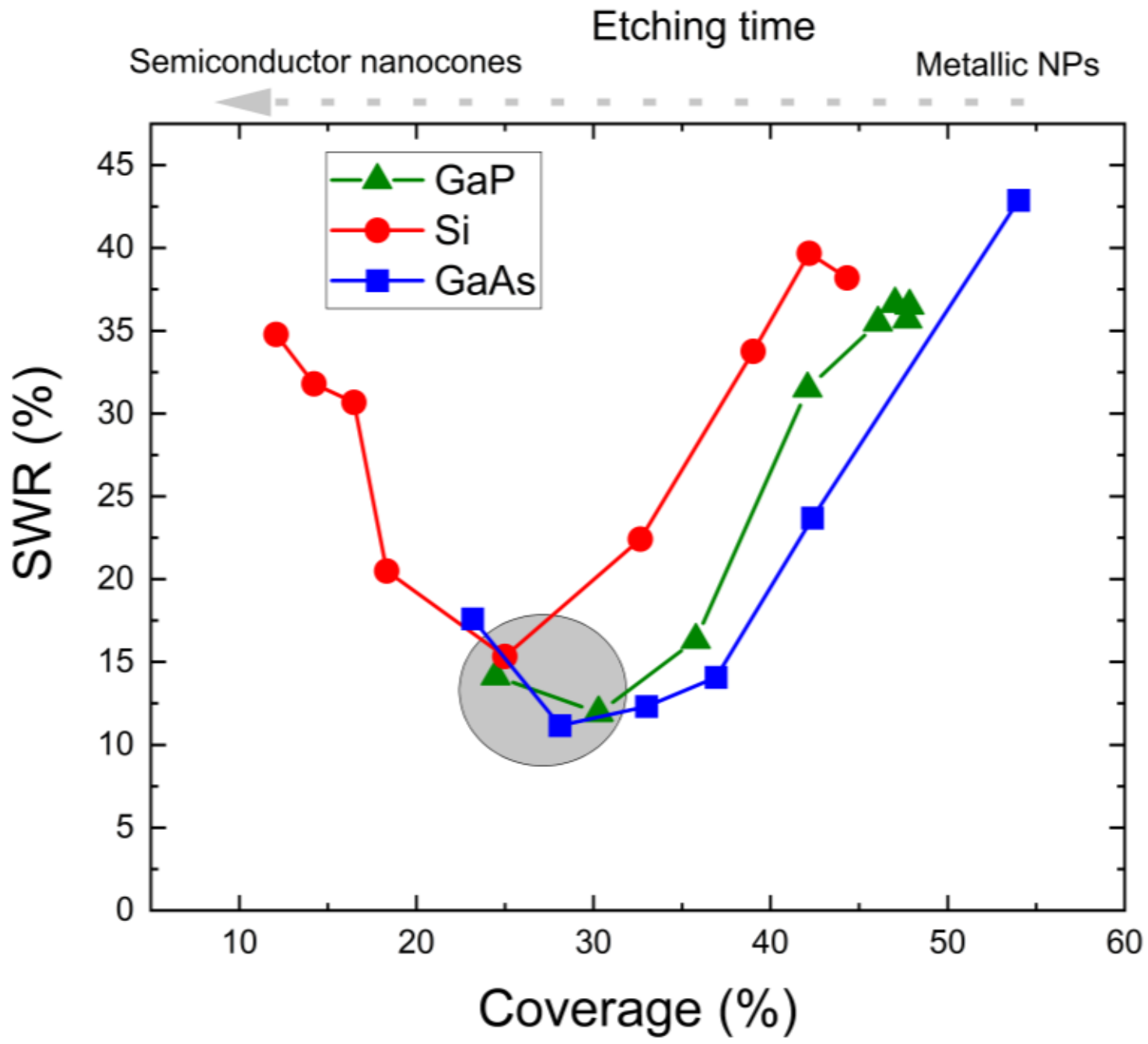

*Figure S17. Sun-weighted reflectance (SWR) as a function of nanostructure surface coverage for GaAs, GaP and Si. Coverage values were extracted from the SEM time series shown in Figs. S5-S7. The graph should be visualized from right to left side as the etching evolves. The reflectance minima for the three materials occur at a comparable coverage of approximately 25-30%, independently of the etching time and residual Ga content.*

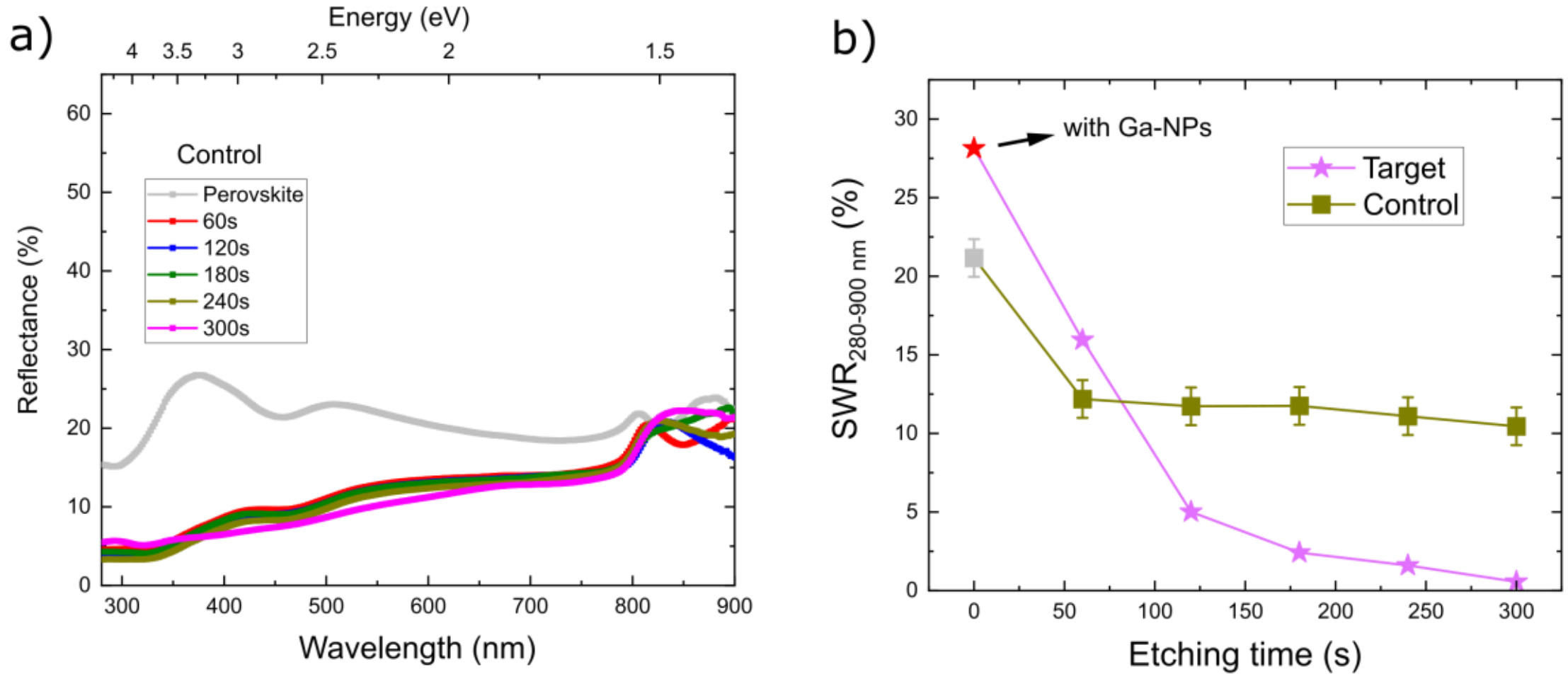

*Figure S18. Reflectance evolution of directly etched perovskite control films. (a) Spectral reflectance measured after successive Ar ICP-RIE times. (b) Corresponding sun-weighted reflectance as a function of etching time. Direct plasma treatment produces only a limited reflectance reduction, which saturates after the initial etching stage and is accompanied by the chemical and optical degradation identified by Raman and absorbance measurements.*